\documentclass[11pt,letterpaper]{article}

\RequirePackage[OT1]{fontenc}
\RequirePackage{amsthm,amsmath}
\RequirePackage[numbers]{natbib}
\RequirePackage[colorlinks,linkcolor=blue,citecolor=blue,
                urlcolor=blue]{hyperref}
\usepackage[margin=3.5cm]{geometry}
\usepackage{amssymb}
\usepackage{authblk}

\usepackage{array}
\usepackage{graphicx}
\usepackage{subfigure}
\usepackage{bm}
\usepackage{siunitx}

\newcommand{\eps}{\varepsilon}
\newcommand{\reals}{\mathbb{R}}

\DeclareMathOperator*{\argmin}{argmin}
\newcommand*{\bdot}{\raisebox{-0.45ex}{\scalebox{1.5}{$\cdot$}}}

\newcommand{\cvar}{t}
\newcommand{\Kfun}{\mathbb{K}}
\newcommand{\fvar}{v}
\newcommand{\fvarp}{u}
\newcommand{\Hil}{\mathcal{H}}
\newcommand{\Kmat}{K}
\newcommand{\xv}{x}
\newcommand{\xdum}{z}
\newcommand{\timecol}{\mathcal{T}}
\newcommand{\timecolp}{\timecol'}
\newcommand{\Xb}{\bm{X}}
\newcommand{\coa}{\alpha}
\newcommand{\cob}{\beta}
\newcommand{\coc}{\gamma}
\newcommand{\coC}{w}
\newcommand{\Gram}{G}
\newcommand{\xvt}{\widetilde{\xv}}
\newcommand{\coah}{\widehat{\coa}}
\newcommand{\cobh}{\widehat{\cob}}
\newcommand{\coch}{\widehat{\coc}}
\newcommand{\regr}{\rho}
\newcommand{\scr}{s}
\newcommand{\tp}{t'}
\newcommand{\Np}{N'}

\newtheorem{rem}{Remark}

\title{\Large Structured functional regression models \\for high-dimensional
    spatial spectroscopy data} 

\author[]{Arash  A. Amini}
\author[]{Elizaveta Levina}
\author[]{Kerby A. Shedden}
    
\affil[]{University of Michigan}

\begin{document}

\maketitle
    

\begin{abstract}
  Modeling and analysis of spectroscopy data is an active area of
  research with applications to chemistry and biology.  This paper
  focuses on analyzing Raman spectra obtained from a bone fracture
  healing experiment, although the functional regression model for
  predicting a scalar response from high-dimensional tensors can be
  applied to any spectroscopy data. The regression model is built on a
  sparse functional representation of the spectra, and accommodates
  multiple spatial dimensions. We apply our models to the task of
  predicting bone-mineral-density (BMD), an important indicator of
  fracture healing, from Raman spectra, in both the in vivo and ex
  vivo settings of the bone fracture healing experiment.  To
  illustrate the general applicability of the method, we also use it
  to predict lipoprotein concentrations from spectra obtained by
  nuclear magnetic resonance (NMR) spectroscopy.

\end{abstract}

\section{Introduction}
Spectroscopy is a type of chemical analysis that infers the
composition of a specimen by using information revealed when it
interacts with radiated energy.  Spectroscopic imaging extends the
utility of traditional spectroscopy by using spectra obtained for
multiple spatial locations within a specimen to map the spatial
distributions of its chemical constituents.  Since some forms of
spectroscopy can be carried out non-invasively and without exposing
research subjects to hazards such as ionizing radiation or chemical
tracers, potential uses of spectroscopy in medical imaging and
diagnostic testing are being actively explored.

Raman spectroscopy is a type of spectroscopy that is based on
inelastic light scattering~\cite{Hanlon2000}.  Like other more
familiar forms of spectroscopy such as infrared (IR) and nuclear
magnetic resonance (NMR/SMR) spectroscopy, Raman spectroscopy produces
a spectrum that expresses the intensity of Raman scattering at each wavelength. This spectrum encodes the chemical structure of the
specimen. Raman spectroscopy uses laser light as an energy source and
is relatively inexpensive compared to MR spectroscopy.  However, the
signal strength for Raman spectroscopy is low, and for human or animal
subjects images can only be
acquired at the depth to which the laser light can penetrate the body.
Raman imaging can be used to image through thin layers of skin and
connective tissue.  For example, Raman maps of bone composition in the
human wrist area, or in limbs of small animals, can be obtained
non-invasively using transcutaneous Raman imaging~\cite{Matousek2006}.

Both Raman and MR spectra exhibit a linearity property that allows the
chemical composition of a complex mixture of pure components to be
resolved.  The measured spectrum of a sample composed of multiple
chemical constituents is approximately equal to the weighted pointwise
sum of the pure component spectra, weighted by the abundance of each
component within the specimen.  An extensive literature in statistics
and
chemometrics~\cite{Chew2002,Widjaja2003,Martin2004,Pauca2006,Vrabie2007}
has focused on techniques for resolving the pure component spectra,
and for quantifying the abundance of each constituent within a
specimen.  Much of this work focuses on analyzing collections of
individual spectra, rather than on spatial maps.

Many promising applications of spectroscopy involve assesment of
biological specimens for diagnostic purposes.  Raman maps of bone
composition obtained at the wrist may provide a useful imaging
biomarker for early detection of bone diseases such as osteoarthritis,
or to non-invasively monitor bone healing following a
fracture~\cite{Tchanque-Fossuo2013, Matousek2006}.  In another
context, NMR spectroscopy has been used to quantify the lipoprotein
fractions in plasma samples~\cite{Dyrby2005}.  These can both be
viewed as prediction problems, in which the information to be used for
prediction is either a single spectrum or a collection of spectra.
This prediction problem can be approached by applying regression
techniques to relate the clinical outcomes to the raw spectra, using a
training set of labeled data.  Alternatively, a more mechanistic
approach can be used in which chemical signatures known to be related
to the variable of interest, such as mineral components in the spectrum of
a healing fracture, are used to predict the outcome.  This approach is
often guided by external information in the form of libraries of pure
bone and tissue spectra, obtained ex vivo from similar specimens. For
recent advances in this line of work, we refer to~\cite{Maher2013}.

The motivation for the models to be
discussed in this paper is a more challenging task of estimating bone composition from Raman
data collected around a larger body part, e.g., around the leg of a
rat. The difficulty in this case is that the non-negligible amount of tissue surrounding the
bone contributes components to the spectrum. Due to this mixing of
signals from bone and tissue, and the weakness of the Raman signal, it
is not yet known whether bone composition can be accurately predicted from
such in vivo Raman readings, and we are not aware of existing successful
attempts to predict an objective measure of bone composition solely
from in vivo Raman data.  

In this study, we focus on predicting
bone-mineral-density (BMD) obtained by micro-CT scans as a ``gold
standard'' for bone composition and fracture healing
(see more on the study set-up in  Section~\ref{sec:exper:setup}.  The BMD is an important predictor
of fracture healing,  and micro-CT scans are a very
accurage way to estimate the BMD, but they are expensive and involve
higher doses of radiation.    

We propose a regularized regression approach for using spectroscopic
images for prediction.  Framing the analysis in a predictive context,
rather than as separate feature extraction and modeling stages, leads
to the discovery of image features that are specifically targetted to
the predictive task.  However this leads to a very high dimensional
regression (e.g.\ the Raman imaging study discussed below has $544
\times 10 \times 5$ features and 37 independent samples), so some form
of regularization is essential.  We propose the use of several
regularizers that respect both the functional forms of individual
spectra, and the way in which the multiple observed spectra for each
subject are related.

The paper is organized as follows. In Section~\ref{sec:exper:setup},
we outline the setup for a fracture healing experiment, in which
spatial in vivo Raman data were obtained from a collection of rats,
with the aim of predicting progression of healing. Since our models
are mainly motivated by the data from this experiment, we will discuss
the setup and the nature of the data in some detail. However, the
models are general and can be applied in other spectroscopy
settings.    As a further example, we also consider
the ex vivo Raman microscopy data from the same fracture healing
experiment. The ex vivo data have higher signal-to-noise than the in
vivo data and are expected to provide much more accurate estimates of
the BMD, however, they lack the rich spatial structure of the in
vivo data, and also of course obtaining ex vivo data requires
sacrificing the animals and is of no use as a diagnostic tool.   
Another type of spectroscopy where our models can be beneficial is
NMR, and we briefly demonstrate their application to a NMR experiment.

Section~\ref{sec:models:methods} is devoted to the discussion of our
models. We first propose, in Section~\ref{sec:func:rep}, a
representation of the data which takes into account its functional
nature. The representation, based on a functional version of the
Lasso~\cite{Tibshirani1996}, simultaneously achieves denoising and
compression. Our main regression model, discussed in
Section~\ref{sec:reg:model}, builds on the functional representation
and takes into account the tensor aspect of the data. In
Section~\ref{sec:results}, we report some empirical results regarding
application of the models to the three datasets described in 
Section~\ref{sec:exper:setup}. The paper concludes, in
Section~\ref{sec:discuss} with a discussion of present shortcomings
and possible extensions of this work.

\section{Experimental data}\label{sec:exper:setup}

\subsection{In vivo Raman data}\label{sec:invivo:Raman:data}
The models in this paper are primarily motivated by the Raman data
from a bone fracture healing experiment. The study was conducted on
$30$ rats.   Each rat underwent a
surgical procedure to induce a small defect in one of its tibias,
removing a thin slice of bone and fixing the bone to a metal plate so
that it can heal back. The rats were then monitored for an eight-week
period, at two-week intervals, starting from week 2.
 Six rats were sampled at all the four time points, while
the rest were only sampled at a single time point (either at week 2,
4, 6 or 8) and then sacrificed to collect ex vivo Raman and micro-CT
data. We have
discarded week 2 data, due to equipment calibration issues that were
resolved later.  In total, $n=37$ usable rat-week samples are available for the analysis.  

\begin{figure}
\centering
\includegraphics[scale=0.5]{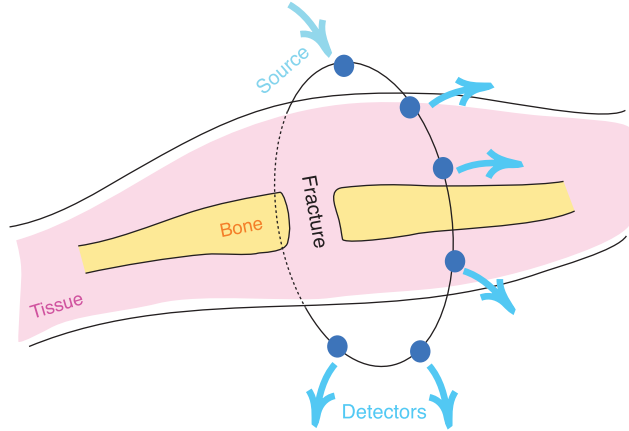}
\caption{A schematic of the ring-shaped apparatus used for obtaining
  the in vivo Raman data from across the leg of a rat.}
\label{fig:ring:schematic}
\end{figure}

In order to collect the Raman spectra, a ring-shaped apparatus was
devised, a schematic of which is illustrated in
Figure~\ref{fig:ring:schematic}. The ring has $d=10$ holes around its
circumference, such that either an illumination or a detection fiber
can be inserted into each hole.  The illumination (source) fiber emits
laser light, and the detection fibers capture the resulting scattered
light. At any given time, only one illumination fiber is used, and the
remaining holes contain detection fibers.


To obtain each Raman measurement, the ring is placed around the leg of
the rat, aligned with the defect, the source is placed in some
position, the scattered light is collected, and the process is
repeated by moving the source to a different position. In total, the
source assumes $p = 5$ positions around the periphery of the ring.

The spectra produced at each of the detector positions are waveforms
expressing the intensity of incident scattered light for wavenumbers
approximately in the range from $\SI{954}{cm^{-1}}$ to $\SI{1800}{cm^{-1}}$. 
We have
truncated the upper end of the spectrum around $
\SI{1700}{cm^{-1}}$ point, since there is thought to be little
relevant information beyond that point. The total number of
wavenumbers, after truncation, is $N = 544$.  

\begin{figure}[t]
  \centering
  \includegraphics[scale=0.27]{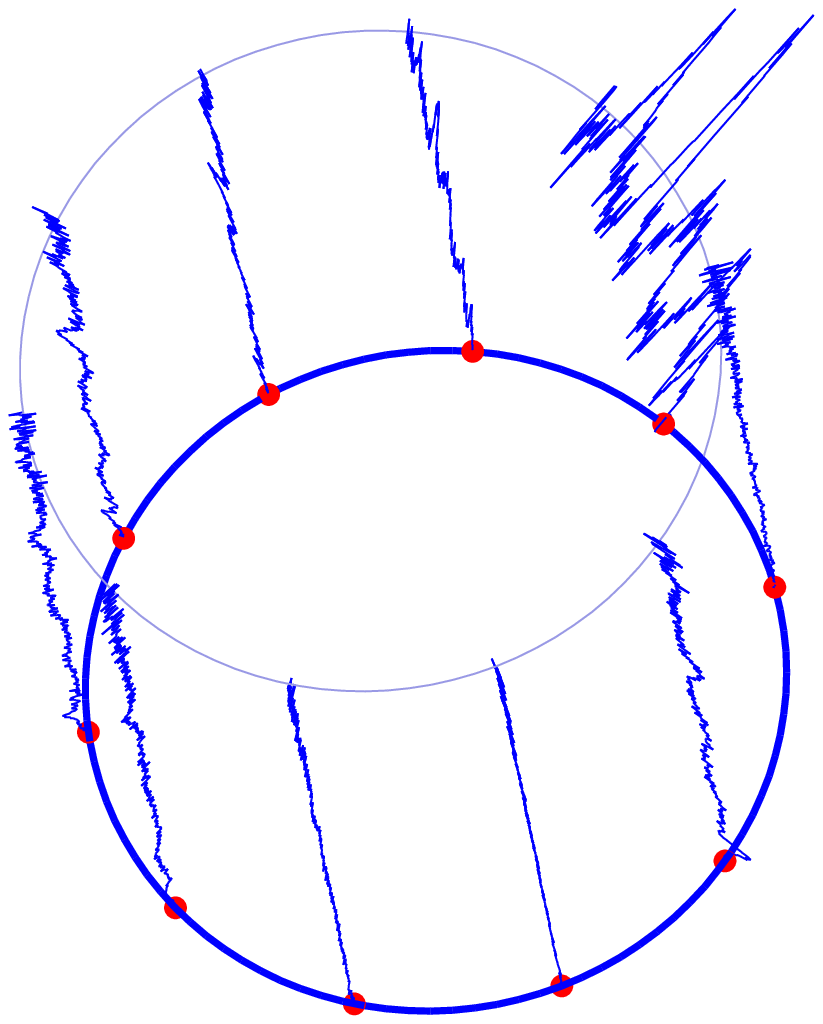}
  \includegraphics[scale=0.27]{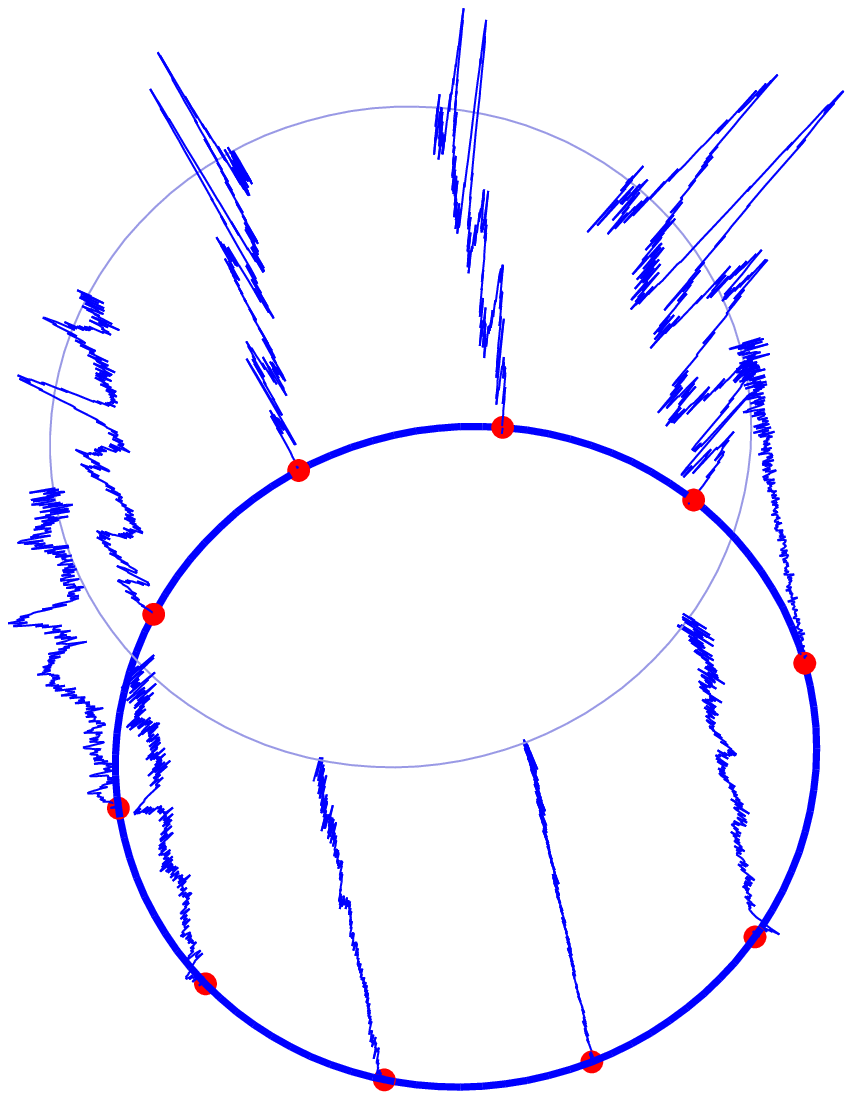}
  \includegraphics[scale=0.27]{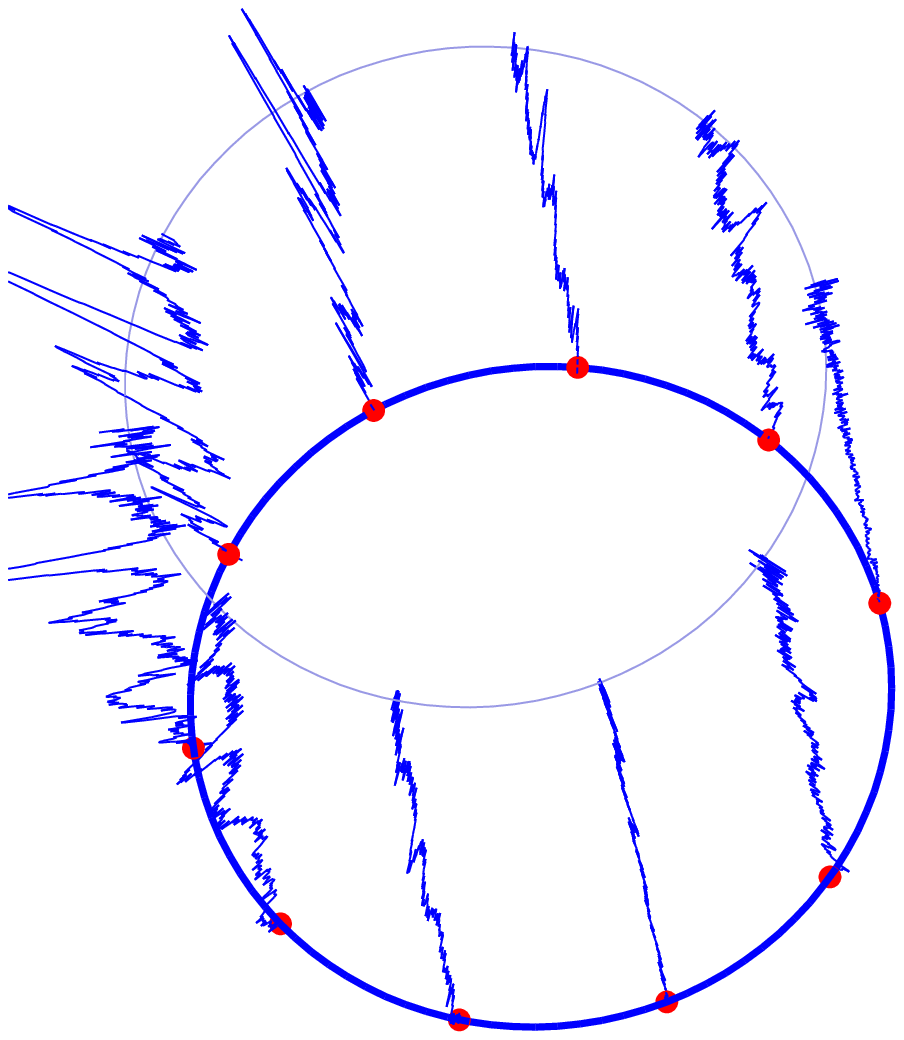}
  \includegraphics[scale=0.27]{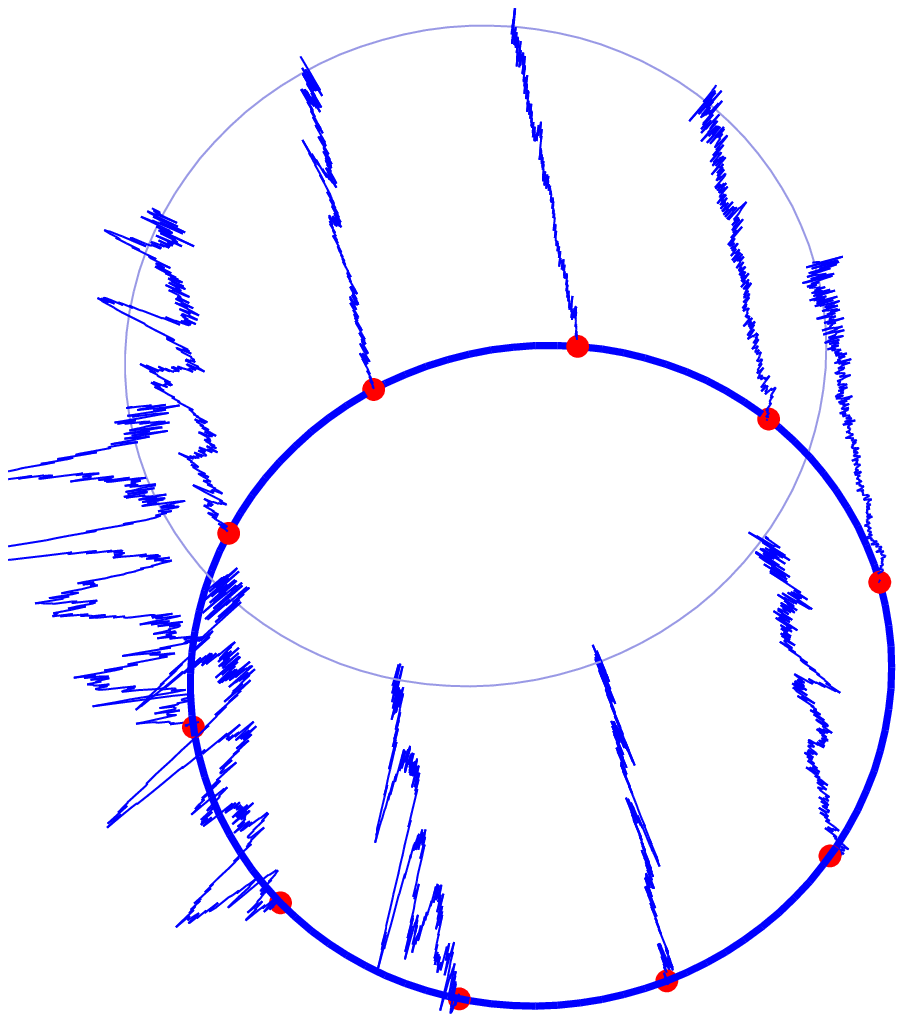}
  \includegraphics[scale=0.27]{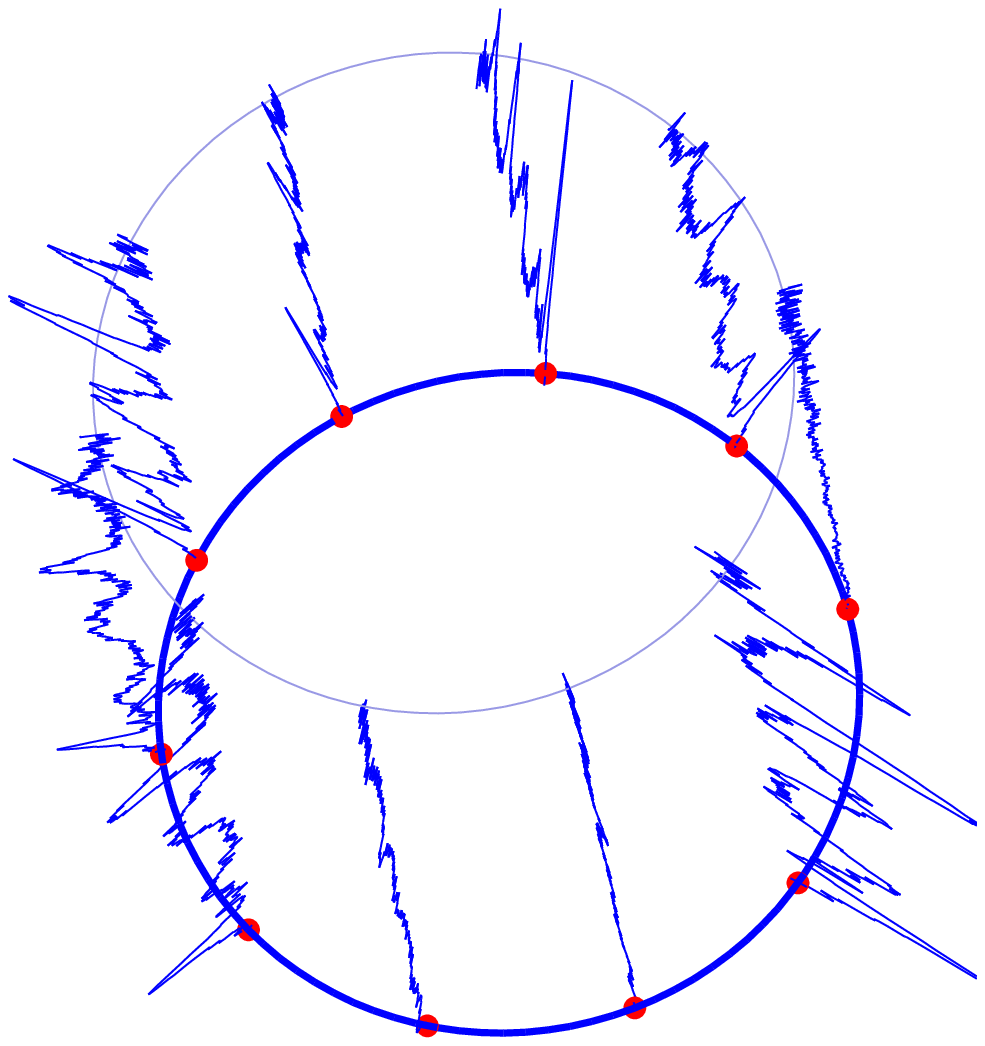}
  \caption{\small Ring view of in vivo Raman tensor for each of the $p=5$ source
    positions. Each cylinder represents all the spectra collected
    for a particular source. The dimension that goes into the page
    represents the wavenumber. The dots around the ring correspond
    (roughly) to detector positions (one of which is also a source
    position in each case.)}
  \label{fig:ring-view}
\end{figure}

\begin{figure}[t]
  \centering
  \includegraphics[scale=0.4]{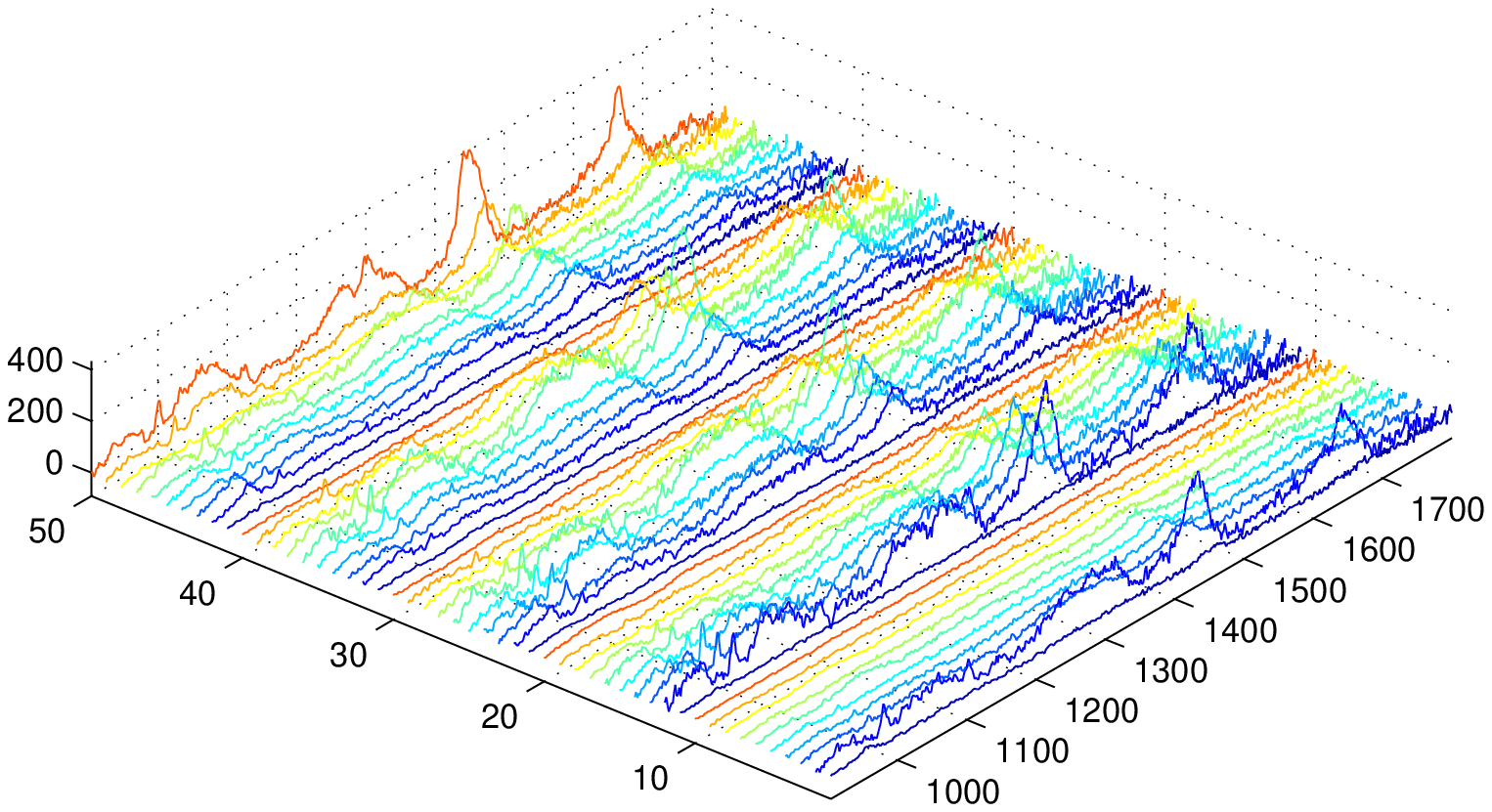}
  \includegraphics[scale=0.4]{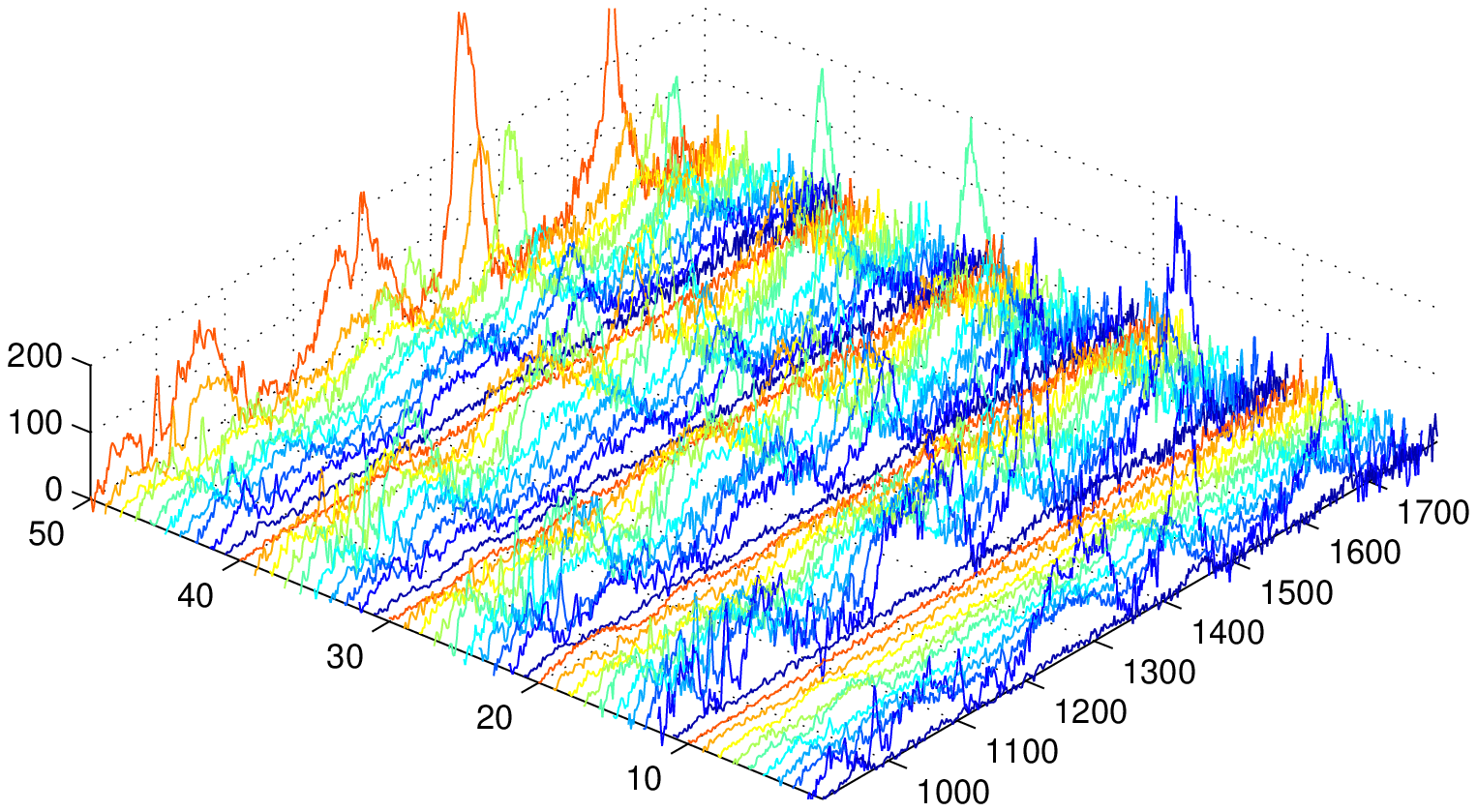}
  \includegraphics[scale=0.4]{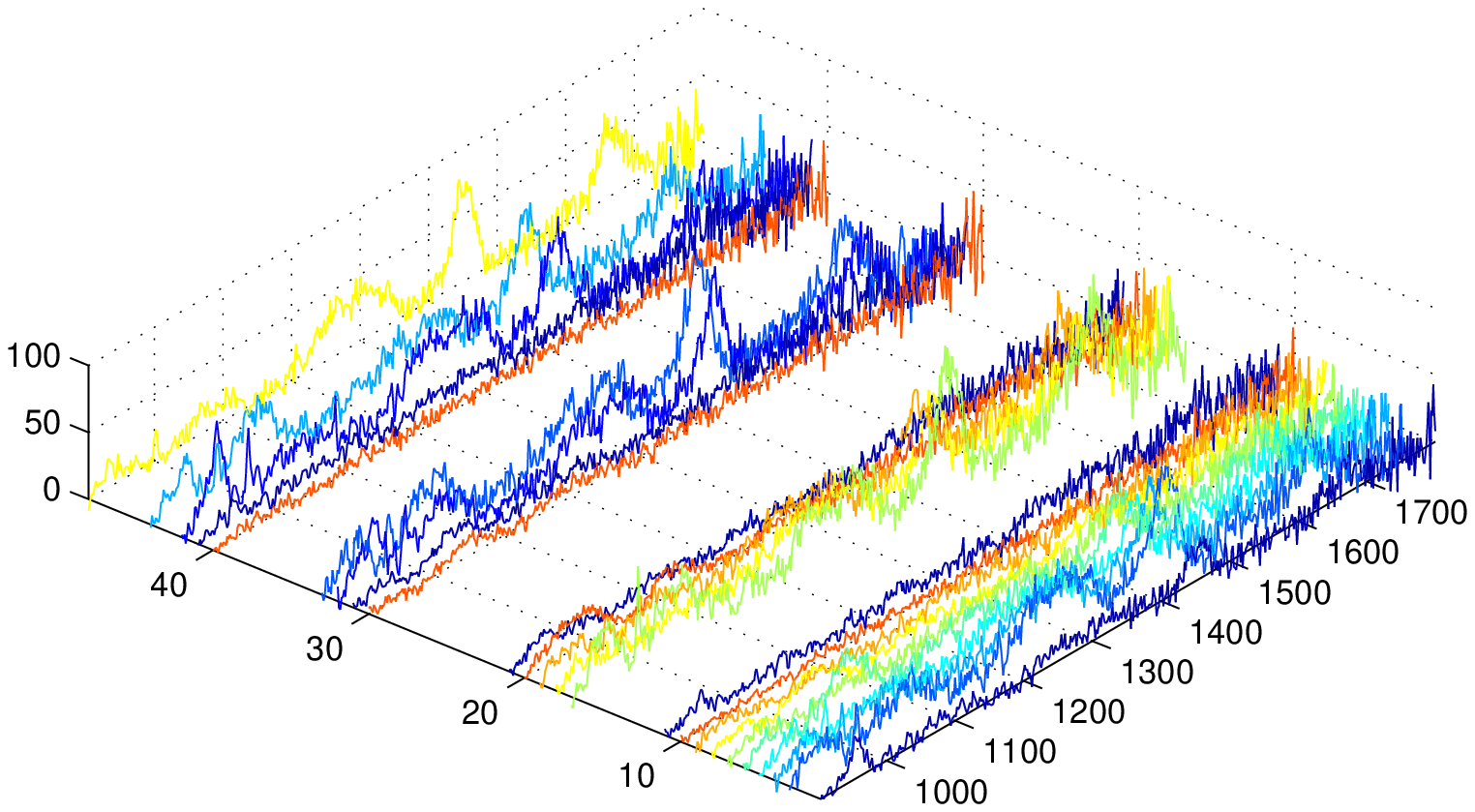}
  \includegraphics[scale=.4]{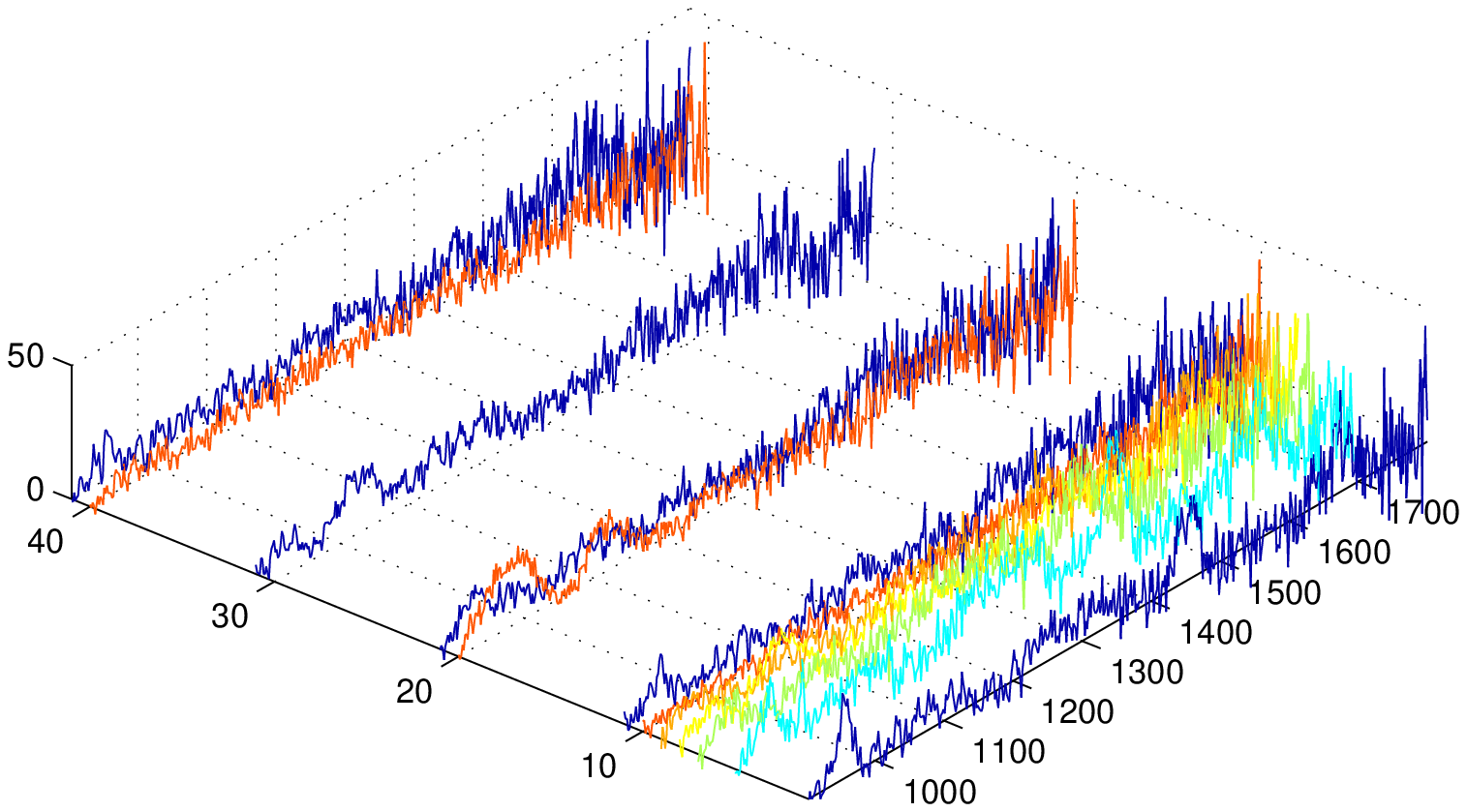}
  \caption{\small Flattened view of the in vivo Raman tensor. The same data is
    shown on four different scales of $y$-axis. In each plot, the spectra with
    peak amplitudes exceeding the scale are not shown. }
  \label{fig:flat-view}
\end{figure}

To summarize, the data for each rat obtained on a single measurement
occasion can be viewed as a $544 \times 10 \times 5$ array, with the
first dimension corresponding to wavenumbers, the second to the
receiver position, and the third to the source position.
Figure~\ref{fig:ring-view} visualizes these data by placing
each detector waveform at the corresponding position around the
ring. Note that as the source rotates, the location of the highest
amplitude detector rotates too. In general, the closer the detector is
to the source, the larger is the amplitude of its
waveform. Figure~\ref{fig:flat-view} is another visualization of a
single measurement, where the source and detector dimensions are
stacked to obtain a $544 \times 50$ matrix. The four different plots in
Figure~\ref{fig:flat-view} show the same measurement at different
scales. In each plot, the spectra whose amplitude exceeds the scale
are omitted from the plot.

Figure~\ref{fig:flat-view} clearly shows the need for normalization. Different
source-detector combinations produce spectra of highly different
amplitudes, which cannot be explained by the relative source/detector
positions. Also, though not visible in the figure, there is
variability betweeen different measurement times. That is, on
different occasions, the same source-detector combination might
exhibit different gains (leading to a different overall amplitude each
time). It is clear that one needs some form of normalization to
combine spectra of such variable scale. Normalization is addressed in
our model by assigning a coefficient to each source and each detector,
so that a proper normalization is learned from the data. We also
employ some simple a priori normalization before applying the model
as a form of pre-processing. At the experimental level, normalization
can be addressed by embedding some form of reference in the sample, to
be used later for calibration. This is often done by using, for
example, a
polymer fiber which produces a strong Raman signal for the polymer
itself. The drawback is some interference with Raman signal of
interest from bone and tissue, and this method was not used in our
study.  

Another issue seen especially in Figure~\ref{fig:flat-view} is the
presence of some weak waveforms (cf. the bottom right plot) that do
not exhibit the overall pattern of a Raman spectrum. These might be due
to the detector being out of position (not properly touching the
tissue) or the source-detector positions being far apart, leading to a
low signal-to-noise ratio. It is desirable to have an automatic
procedure for discarding these \emph{noise spectra}.  Our model
addresses this issue, to some extent, by picking the relevant
source-detector combinations. This is another consequence of assigning
a coefficient to each source and detector, and estimating them based on
the data. A third issue illustrated by Figure~\ref{fig:flat-view}, is
the noisy nature of Raman data, which is especially evident at lower
scales, suggesting that some form of denoising might be helpful. This
will be automatically handled by our functional representation, to be
discussed in Section~\ref{sec:func:rep}

Our main goal for the Raman data from this experiment is to test its
ability to 
predict well-established biomarkers of fracture healing which can be
obtained by the more costly micro-CT approach. The mirco-CT data is
obtained ex vivo, and it can produce measures such as bone mineral
content (BMC), bone mineral density (BMD), tissue mineral content
(TMC), and tissue mineral density (TMD). Although the micro-CT data
can produce a high-resolution spatial map of these quantities inside
the specimen, we only had access to an average value for each rat
within the region of interest. The different markers are highly
correlated and we have elected to consider only the BMD
sequence. Thus, our response is a sequence of $n = 37$   scalar BMD
measurements. In Section~\ref{sec:reg:model}, we will discuss our
regression model relating the Raman tensor to the BMD response.

\subsection{Ex vivo Raman data}\label{sec:exvivo:Raman:data}
The ex vivo Raman dataset corresponds to the same experiment described
in Section~\ref{sec:invivo:Raman:data}. The data are Raman spectral
maps of cross sections of bone obtained ex vivo after the rats are
sacrificed. A cross-section of pure bone (separated from tissue) is
placed under a microscope fitted with excitation laser and detection
devices to collect Raman spectra, at precise locations inside the
cross-section.

The specific locations at which the Raman spectra were collected
differed by rat, based on bone morphology. The number of measurement
locations varied from 3 to 7 (with up to 3 sub-locations for each). We
averaged the spectra within rats to obtain a single average spectrum
per rat.  Since the BMD is also an average value, this can be viewed
as a prediction of spatially averaged BMD from spatially averaged
Raman spectra.  In total, there were 23 rats measured at $806$
wavenumbers.

\subsection{NMR data}\label{sec:nmr:data}
To further explore the effectiveness of our approach, we also
considered a dataset of 2D diffusion-edited H NMR spectra. This
dataset is extensively studied in~\cite{Dyrby2005}. It is also used
by~\cite{Zhao2013} to illustrate their matrix regression
approach, discussed below in Remark~\ref{rem:compare:with:Zhao} in
Section~\ref{sec:reg:model}).  The dataset contains NMR spectra and
lipoprotein concentrations for $25$ human subjects. The concentrations
of cholesterol and triglyceride were obtained by ultracentrifugation,
for various fractions and subfractions in terms of lipoprotein
density. The primary fractions of interest are very low, low,
intermediate, and high density lipoproteins, abbreviated as VLDL, LDL,
IDL and HDL. A total of $32$ concentration levels are reported, from
which we have used the $4$th  variable \verb+`CH_V2'+, following
\cite{Zhao2013}, which stands for Very
low-density CHolesterol, subfraction 2). For this variable, the
concentrations were missing for $5$ subjects; hence, we take the
response vector to be the $n=20$ (scalar) recorded concentration levels, and discard
the spectra corresponding to missing responses.

\begin{figure}
  \centering
   \includegraphics[scale=.7]{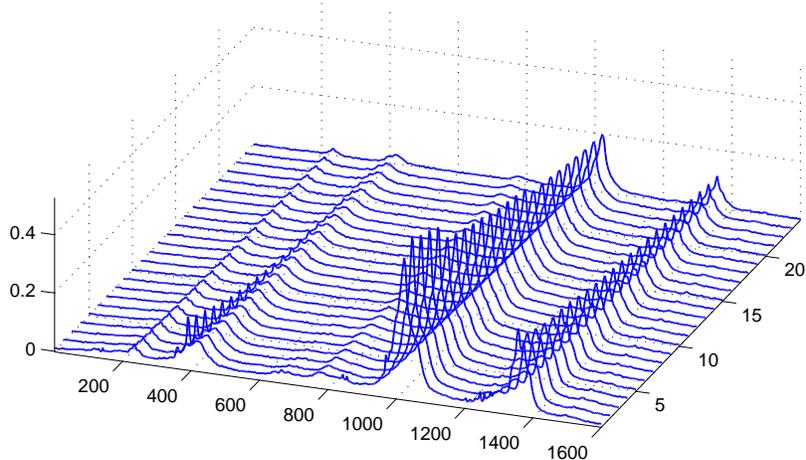}
   \caption{\small A typical example of a NMR spectrum. The spectral
     dimension runs to $1600$. }
   \label{fig:nmr:spec}
\end{figure}

The NMR spectra are measured as intensity for each chemical shift (in
the range 2.5--0.6ppm); see~\cite{Dyrby2005} for details. For each
subject, a 2-D spectrum of dimensions $24 \times 1600$ is produced,
where the second dimension is the spectral
one. The first dimension corresponds to 24 steps of  gradient pulse strength. Figure~\ref{fig:nmr:spec} illustrates a typical example. Note
that in this plot (and elsewhere) the spectral range is indexed
sequentially from $1$; it does not show the true value chemical shift, since
we did not have access to the exact value.  In terms of
dimension, this dataset lies between the in vivo and ex vivo
fracture healing Raman datasets. It has both a spatial and a spectral
dimension, making it a good fit for our modeling approach. Since the
spectra are smooth along the spectral dimension, we have sub-sampled
them at a rate $1/3$ as a preprocessing step to reduce the dimension
of each spectrum to $534 \times 24$.


\section{Models and methods}\label{sec:models:methods}

In this section, we present our models for spectroscopy data, in
general notation, but tailored to the in vivo Raman dataset of
Section~\ref{sec:invivo:Raman:data}.  We consider data that have one
continuous (functional) dimension and several discrete (spatial)
dimensions. In order to describe the model, we consider the case where
there are two discrete dimensions and a single continuous one,
although the results can be easily generalized.  To be specific, the
data is a collection
\begin{align*}
\big\{X_{ij}^{k}(\cdot): \, (i,j,k) \in [d]\times [p] \times [n]
\big\}
\end{align*}
of functions, recorded at spatial positions $(i,j)$. Here,
we are using the notation $[p] := \{1,\dots,p\}$ and similarly for
$[d]$ and $[n]$. Index $k$ is used to enumerate the statistical
samples. In other words,
\begin{align*}
  \{X_{ij}^{1}(\cdot), \ (i,j)  \in [d]\times [p] \}, \; \{X_{ij}^{2}(\cdot) , \ (i,j)  \in [d]\times [p] \}, \;
  \dots, \{X_{ij}^{n}(\cdot) , \ (i,j)  \in [d]\times [p] \}
\end{align*}
are assumed to be i.i.d. samples in an experiment.  We will
use $\cvar$ to denote the continuous index, and assume that each
function $\cvar \mapsto X_{ij}^k(\cvar)$ is observed over some
interval $T \subset \reals$. Moreover, we refer to the continuous
domain (i.e, $T$) as the \emph{spectral domain} and to its elements
(i.e., $t \in T$) as \emph{wavenumbers}, based on our main application
to in vivo Raman spectra, though the model is general. Similarly, we may refer to indices $i$ and $j$ as source and
detector positions, respectively.

We consider two analysis goals involving this type of data. The first
is that of finding an efficient (i.e., compact) and informative
representation of the data. The second is that of using the relatively
high-dimensional dataset $\{X_{ij}^k(t)\}$ as covariates in a
regression problem to predict a response vector $\{y^k\}$. These goals
are related, since obtaining a compact representation of the data
greatly facilitates the regression analysis.

\subsection{Functional representation}\label{sec:func:rep}

In order to obtain a compact representation of the data, we take each
function $X_{ij}^k(\cdot)$ to lie in a reproducing kernel Hilbert
space (RKHS)~\cite{Berlinet2004}, generated by some kernel function
$\Kfun: T \times T \to \reals_+$. Usually, the functions
$\{X_{ij}^k(\cdot)\}$ are only observed at a discrete set of points
$\timecol := \{t_1,t_2,\dots,t_N\}$. We additionally assume that each
$X_{ij}^k(\cdot)$ can be well approximated by a finite linear
combination of the kernel functions anchored at points of
$\timecol$. That is,
\begin{align}\label{eq:kern:rep}
  X_{ij}^k(\cvar) \approx \sum_{\fvar = 1}^N \xv_{ij\fvar}^k\;
  \Kfun(t,t_\fvar), \quad t \in T.
\end{align}
This assumption simplifies the subsequent derivations and is motivated
(and somewhat justified) by the representer
theorem~\cite{Scholkopf2001}.  The kernel function
$\Kfun$ can be taken to be any valid kernel (positive semi-definite,
symmetric bivariate function), though our main focus will be on the
Lorentzian
\begin{align*}
  \Kfun(t,s) := \frac{1}{1 + (\frac{t-s}{W})^2},
\end{align*}
where $W$ is a bandwidth parameter.  Empirically, this
kernel provides a good model for pure spectra, and is also justified
by physical considerations~\cite{Meier2005}. Another restriction that
we impose is for the coefficients $\{\xv_{ij\fvar}^k\}$ to be
nonnegative. This is also in accordance with the physics of how
spectra are formed as a weighted linear combination of 
spectra of pure chemical components, without any cancellations.

Let us fix $(i,j,k)$ for the rest of this section. Based
on~\eqref{eq:kern:rep}, the idea is to turn the collection
$\{X_{ij}^k(\cvar)\}_{\cvar \in T}$ into the set of coefficients $\{ \xv_{ij \fvar}^k\}_{
  \fvar \in [N]}$, which is easier to work with.   To achieve a
compact representation, 
we impose a sparsity constraint on the vector $\xv_{ij \bdot}^k:=
\{\xv_{ij\fvar}^k\}_{ \fvar \in [N]}$. In other words, we seek a
representation of the form~\eqref{eq:kern:rep} with as few nonzero
coefficients as possible. This can be done by solving the
$\ell_1$-regularized least-squares problem
\begin{align}\label{eq:func:ell1:v1}
\argmin_{\xdum \,\in\, \reals^N_+} \Big\{ \frac12
  \sum_{\fvarp = 1}^N \Big[ X_{ij}^k(t_\fvarp) - \sum_{\fvar = 1}^N
  \xdum_{\fvar}\; \Kfun(t_\fvarp,t_\fvar)\Big]^2 + \lambda_\Hil
  \sum_{\fvarp,\fvar = 1}^N \xdum_\fvarp \xdum_\fvar \Kfun(t_\fvarp,t_\fvar) +
  \lambda_1 \sum_{\fvar=1}^N |\xdum_\fvar| \Big\}.
\end{align}
Here, $\reals^N_+$ is the set of $N$-vectors with
nonnegative components.  We note that the term $\sum_{\fvarp,\fvar}
\xdum_\fvarp \xdum_\fvar \Kfun(t_\fvarp,t_\fvar)$ is the RKHS norm of
the function $\sum_\fvar \xdum_\fvar \Kfun(\cdot,t_\fvar)$.  When the
RKHS norm measures smoothness of the function, regularizing by this term leads to smoother solutions.

\begin{figure}\label{fig:kern:rep}
  \centering
  \includegraphics[scale=0.65]{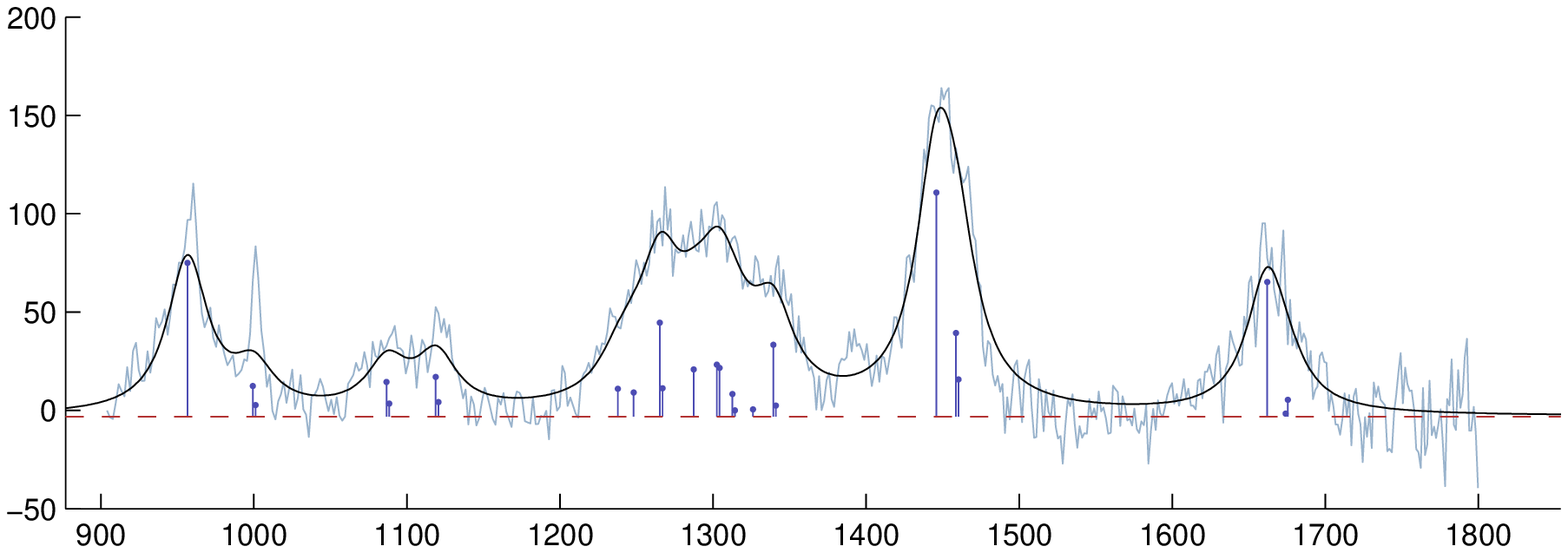}
  \includegraphics[scale=0.65]{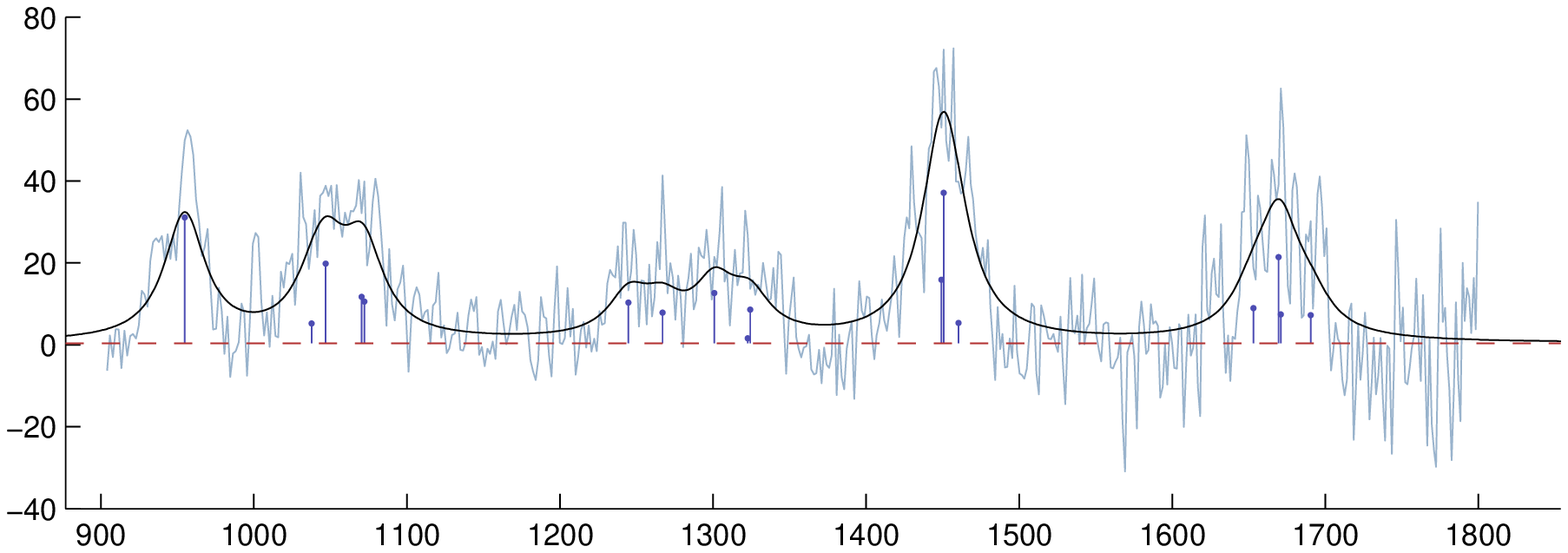}
  \caption{\small Examples of fitting model~\eqref{eq:kern:rep} to Raman
    spectra. The fitted spectra are shown in black; the estimated
    non-zero coefficients $\{\xv_{ij\fvar}^k\}$ are shown as vertical
    lines, with height representing their magnitude.}
  \label{fig:func:rep}
\end{figure}

Let $\Kmat$ be the $N\times N$ matrix with entries
$\Kfun(t_\fvarp,t_\fvar)$, and let $\|\xdum\|_p := \big( \sum_{\fvar =
  1}^N |\xdum_\fvar|^p \big)^{1/p}$ denote the $\ell_p$ norm of $\xdum
= (\xdum_1,\xdum_2,\dots,\xdum_N)$. Moreover, let $\Xb_{ij}^k := \big(
X_{ij}^k(t_\fvarp) \big)_{\fvarp \in [N]}$ so that $\Xb_{ij}^k$ is an
$N$-vector. Then,~\eqref{eq:func:ell1:v1} can be rewritten in the
compact form
\begin{align}\label{eq:func:ell1:v2}
\argmin_{\xdum \,\in\, \reals^N_+}
  \Big\{ \frac12 \| \Xb_{ij}^k - \Kmat \xdum \|_2^2 +
  \lambda_\Hil \, \xdum^T \Kmat \xdum + \lambda_1 \|\xdum\|_1
  \Big\}.
\end{align}
Note that this is a standard convex problem which can be
solved efficiently. Figure~\ref{fig:kern:rep} shows examples of fitted
Raman spectra.

\subsection{Regression model}\label{sec:reg:model}

In this section we devise a model to predict a (one-dimensional)
response vector $\{y^k\}$ based on the observed tensor covariates
$\{X_{ij}^k(t)\}$.  Perhaps the simplest model is to assume that a
rank-one multilinear map relates the covariates to the response, that
is, we consider the following regression model,
\begin{align}\label{eq:reg:mod:cont}
  y^k = \sum_{ij} \int_T \coa_i \cob_j \coC(t) \, X_{ij}^k(t) \,dt + \eps^k
\end{align}
for $k=1,\dots,n$, where $\{\eps^k\}$ are i.i.d. noise
variables. In accordance with~\eqref{eq:kern:rep}, we simplify the
model further by assuming the following representation for $\coC$,
\begin{align}\label{eq:func:weights}
  \coC(t) := \sum_{\fvarp = 1}^N \coc_\fvarp \Kfun(t,t_\fvarp), \quad t
  \in T.
\end{align}
Combining with representation~\eqref{eq:kern:rep} for
$\{X_{ij}^k(t)\}$ and ignoring its approximation error, we arrive at the model
\begin{align*}
  y^k = \sum_{ij \fvarp \fvar} \coa_i \cob_j \coc_\fvarp
  \,\Gram_{\fvarp \fvar} \,\xv_{ij\fvar}^k  + \eps^k
\end{align*}
where $\Gram_{\fvarp \fvar} := \int_{T} \Kfun(t,t_\fvarp)
\Kfun(t,t_\fvar) \,dt$. Note that this is the $L^2(T)$ inner product
of the functions $\Kfun(\cdot,t_\fvarp)$ and
$\Kfun(\cdot,t_\fvar)$. Let $\Gram := (\Gram_{\fvarp \fvar}) \in
\reals^{N \times N}$, and note that this is a Gram matrix. The model
can be written more compactly by defining
\begin{align}\label{eq:reg:mod:disc}
  \xvt_{ij\fvarp}^k := \sum_{\fvar=1}^N \Gram_{\fvarp \fvar}
  \,\xv_{ij\fvar}^k, \quad
\text{so that} \quad
   y^k = \sum_{ij \fvarp } \coa_i \cob_j \coc_\fvarp
  \,\xvt_{ij\fvarp}^k  + \eps^k,
\end{align}
where the summation is over $(i,j,\fvarp) \in
[p]\times[d]\times[N]$. Note that the advantage of this rank-one model
is that it contains $p+d+N$ variables, which in high dimensions is far
less than that of a full linear model with $p d N$ variables.

In order to fit model~\eqref{eq:reg:mod:disc}, we again solve
a regularized least-squares problem. Since $(i,j)$ represent spatial
dimensions in our setting, we do not expect much sparsity in $\coa = (\coa_i)$
and $\cob = (\cob_j)$. Hence we use an $\ell_2$ norm regularizer for
these two. On the other hand, we expect considerable sparsity in the
spectral domain variable $\coc = (\coc_\fvarp)$, hence we regularize by
its $\ell_1$ norm.

The final element of our proposed regularizer is a penalty which tends
to bring the coefficients assigned to nearby wavenumbers closer
together. This is justified if proximity in the spectral domain
signifies similarity, as suspected to be the case for Raman
spectra. This type of regularizer also provides a practical advantage
which is discussed in Section~\ref{sec:practical}. As a measure of
similarity between wavenumbers, we can use the Gram matrix $\Gram$. It
is then natural to use a weighted fused Lasso
penalty~\cite{Tibshirani2005,Chen2012} of the form $\sum_{\fvarp
  \fvar} \Gram_{\fvarp \fvar} |\coc_\fvarp - \coc_\fvar|$.

Putting the pieces together, we solve the following
\begin{align}\label{eq:tensor:rls}
\begin{split}
  (\coah,\cobh,\coch) &=
  \argmin_{\substack{ (\coa,\,\cob,\,\coc):\;
      \\ \coa \,\ge\, 0,\; \cob \,\ge\, 0}}
  \Big\{
  \frac1{2n} \sum_{k=1}^n \Big( y^k - \sum_{ij \fvarp } \coa_i \cob_j \coc_\fvarp
  \,\xvt_{ij\fvarp}^k   \Big)^2
  + \frac{\regr_\coa}{\sqrt{p}} \, \| \coa\|_2 +
  \frac{\regr_\cob}{\sqrt{d}} \, \| \cob\|_2   \\
  & \qquad \qquad \qquad \qquad  \qquad \qquad \qquad
  + \,\frac{\regr_\coc}{N} \, \| \coc\|_1 +
  \frac{\regr_\Gram}{N^2} \, \sum_{\fvarp, \fvar=1}^N
\Gram_{\fvarp\fvar} |\coc_\fvarp - \coc_\fvar| \Big\}
\end{split}
\end{align}
for some positive numbers $\regr_\coa,\regr_\cob,\regr_\coc$
and $\regr_\Gram$. Conditions $\coa \ge 0$ and $\cob \ge 0$ are
interpreted element-wise. They are imposed to remove the
sign-ambiguity which is otherwise present in the model. Note also that
there is a general scale ambiguity in model~\eqref{eq:reg:mod:disc},
but not in~\eqref{eq:tensor:rls} due to the presence of the
regularizers. (By scale ambiguity, we mean that $(\coa,\cob,\coc)$ and
$(c_1\coa,c_2\cob,c_2\coc)$ determine the same multilinear map as long
as $c_1 c_2 c_3 = 1$ holds.)

\begin{rem}\label{rem:compare:with:Zhao}


Enforcing an exact rank-one constraint on the regression function has
been recently proposed in~\cite{Zhao2013} in the context of regression
with matrix covariates. Their model is similar to~\eqref{eq:reg:mod:cont}, without the functional dimension, i.e., $y^k
= \sum_{ij} \coa_i \cob_j X_{ij}^k + \eps^k$. They enforce sparsity on
both sets of coefficients $(\coa)$ and $(\cob)$, using a
multiplicative $\ell_1$ penalty. Our model provides an extension (with
variation) to the tensor case with mixed functional and discrete
dimensions. For us, the coefficients $(\coa)$ and $(\cob)$ have
physical (spatial) interpretations and are not necessarily sparse.
\end{rem}

\begin{rem}
  Enforcing a low-rank assumption on the regression function, but not
  necessarily an exact one, has been studied extensively in recent
  years. The preferred approach to the problem is via nuclear norm
  penalization. We refer to~\cite{Koltchinskii2011,Negahban2011} for
  more details. Empirically, we found that imposing a rank-one
  assumption directly enhances the interpretation by considerably
  reducing the dimensionality. The drawback is the non-convexity of
  the resulting problem (see below) and lack of theoretical
  guarantees.
\end{rem}

\subsection{Practical considerations}\label{sec:practical}

Let us consider some practical issues regarding the models discussed
earlier.

\begin{itemize}

\item The cost function in~\eqref{eq:tensor:rls} is not (jointly)
  convex in $(\coa,\cob,\coc)$, but it is separately convex in each of
  these variables. A standard way to optimize such functions is by
  alternating minimization, fixing two variables at a time and
  minimizing over the other.

\begin{figure}[t]\label{fig:scores}
  \centering
  \includegraphics[scale=0.75]{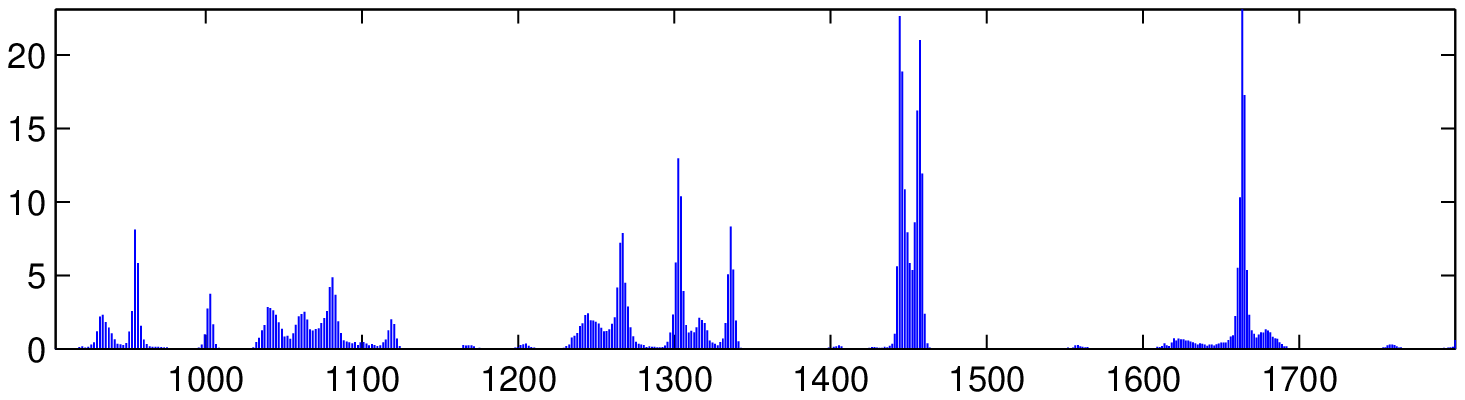}
  \includegraphics[scale=0.75]{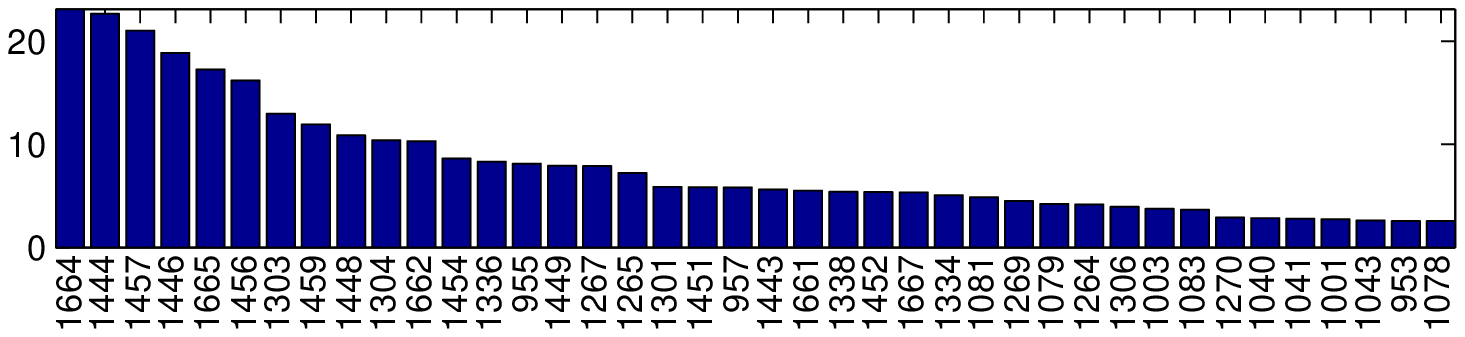}
  \caption{\small Example of the scores obtained based
    on~\eqref{eq:score:def} for a sample of Raman data. The bottom
    plot shows the $40$ wavenmubers with the highest scores.}
\end{figure}

\item In practice, we observe $\{X_{ij}^k(\cdot)\}$ at more points
  than we want to keep in the model. More specifically, we observe the
  functions at a set of points $\timecolp \supset \timecol$, and we
  discard some of the points to obtain the $\timecol$ which is used in
  fitting the model. This is to avoid over-fitting. One approach to
  choose which points to keep is to assign a score to each point in
  $\timecolp = \{ \tp_1,\tp_2\dots,\tp_{\Np}\}$, based on their
  weighted frequency\footnote{We could also work with just the
    frequency, that is $\frac{1}{ p d n} \sum_{ijk} 1{\{
      \xv_{ij\fvar}^k \neq 0 \}}$.}  of appearing in the dataset. For
  example, we can assign the following score to $\tp_\fvar$,

  \begin{align}\label{eq:score:def}
    \scr_\fvar := \frac{1}{ p d n} \sum_{ijk} \xv_{ij\fvar}^k, \quad
    \fvar \in [\Np]
  \end{align}

  where $\{\xv_{ij \fvar}^k\}$ are the coefficients in
  expansion~\eqref{eq:kern:rep} of Section~\ref{sec:reg:model}, and
  the sum runs for $(i,j,k) \in [p]\times [d] \times [n]$. We can then
  keep wavenumbers corresponding to the $N$ largest
  scores. Figure~\ref{fig:scores} shows the scores obtained for the
  example Raman data and the wavenumbers corresponding to $N = 40$
  largest scores. (Here $N' = 544$.)


\item An advantage of the weighted fused Lasso penalty used
  in~\eqref{eq:tensor:rls}, in connection with the previous point, is
  that we do not need to worry about the order of the wavenumbers kept
  in $\timecol$. The Gram matrix $\Gram$ automatically tries to match
  the coefficients of nearby wavenumbers regardless of how they are
  ordered in $\timecol$.

\item It is possible to replace $\ell_2$ norm regularizations for
  $\alpha$ and $\beta$, with constraints $\sum_i \alpha_i = \sum_i
  \beta_i = 1$. This gives $\alpha$ and $\beta$ the natural
  interpretation of being probability vectors. We have not used this
  version here, since it tends to have a similar effect as an $\ell_1$
  regularization (note that $\alpha,\beta \ge 0$), that is, it
  produces sparse solutions. It is more suitable in situations in
  which only few of the spatial positions are suspected to be
  influential, which is not the case for data considered here.

\item In practice, we choose the regularization parameters in
  both~\eqref{eq:func:ell1:v2} and~\eqref{eq:tensor:rls} by
  cross-validation (CV). The scaling of regularization parameters
  in~\eqref{eq:tensor:rls} by $1/\sqrt{p}$, $1/\sqrt{d}$, etc. is to
  keep the corresponding parameters in the same range, and have little
  effect in practice if CV is used. 

\item Although the model~\eqref{eq:reg:mod:cont} is designed to pick
  the correct normalization constants for each source-detector
  combination, through $(\coa)$ and $(\cob)$, in practice some form of
  crude scaling is useful prior to fitting the model, to keep all the
  waveforms on a reasonable scale. We consider two simple approaches,
  each of which can be applied to either $\{\xv_{ij\fvar}^k\}$ or
  $\{\xvt_{ij\fvarp}^k\}$. These are specific to the in vivo Raman
  application, for which $(i,j)$ refers to source-detector position.
  Other approaches might be suitable in different applications.

  The first approach is to normalize, e.g., $\{\xvt_{ij\fvarp}^k\}$ so
  that the maximum amplitude among all the data for a particular
  source-detector combination is $1$. That is, we work with the
  normalized sequence $\{\xvt_{ij\fvarp}^k / (\max_{k,\fvarp}
  \xvt_{ij\fvarp}^k)\}$.
  The other approach is to normalize so that the total energy (in the
  sense of $\ell^2$-norm) corresponding to each source position is
  unit, in each application of the measuring device (i.e., for each
  $k$). More specifically, we look at the rescaled sequence
  $\big\{\xv_{ij \fvar}^k / \sqrt{\sum_{j \fvar} (\xv_{ij \fvar}^k)^2}
  \big\}$. A more elaborate version of this normalization could take
  into account the kernels anchored at each wavenumber and use the
  continuous $L^2$ norm.

  Empirically, the two approaches outlined above produce comparable
  results. For definiteness, the results reported in the
  Section~\ref{sec:results} are based on the maximum amplitude
  normalization (unless otherwise stated).

\end{itemize}


\section{Empirical results}\label{sec:results}

\subsection{In vivo Raman}\label{sec:results:invivo}

We start by applying the models of
Section~\ref{sec:models:methods} to the in vivo Raman data.  Here, the
models are applicable verbatim. To obtain the results, we have
centered the BMD sequence $\{y^k\}$ and normalized so that its maximum
absolute value is equal to $1$, that is, $\frac1n \sum_k y_k = 0$ and
$\max_{k} |y_k| = 1$. The modified Raman tensor
$\{\xvt_{ij\fvarp}^k\}$ is normalized according to the first approach
discussed in Section~\ref{sec:practical}, so as to have the maximum
amplitude of $1$ for every source-detector pair. 

\begin{figure}[t]
    \centering
    \includegraphics[scale=.4]{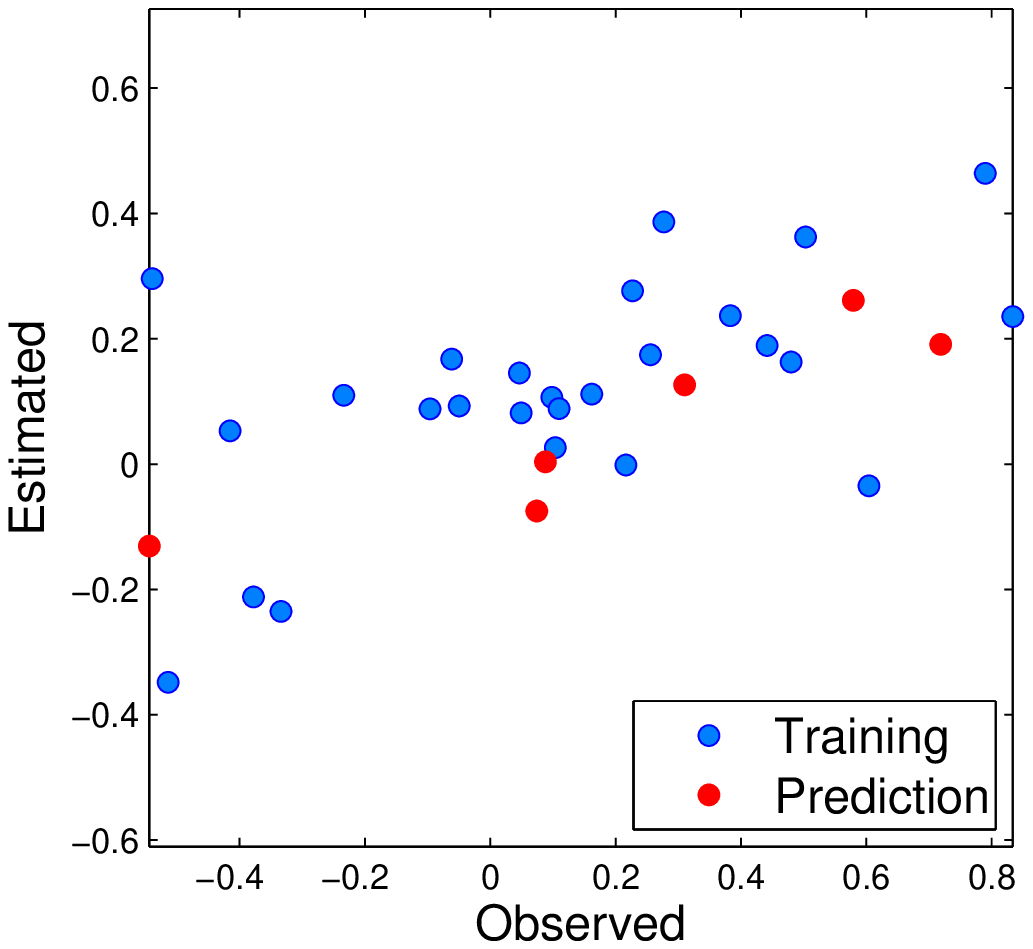}
    \includegraphics[scale=.4]{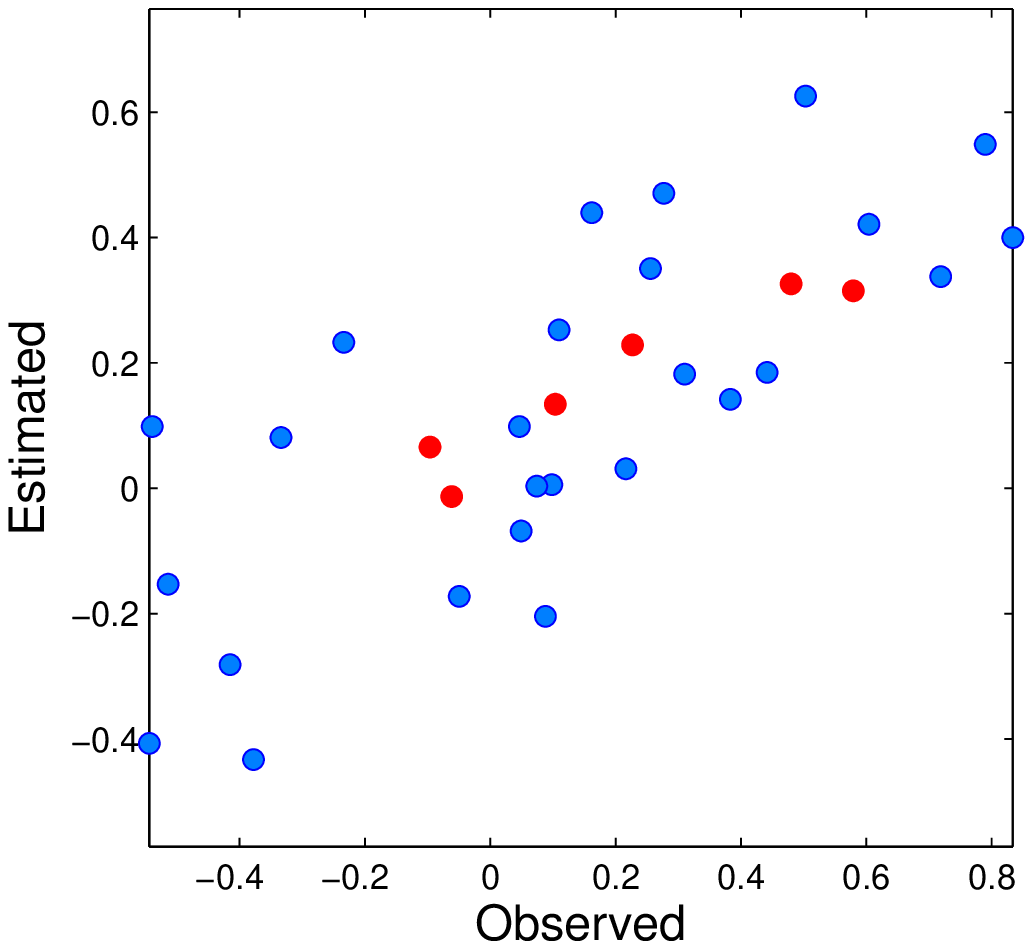}
    \includegraphics[scale=.4]{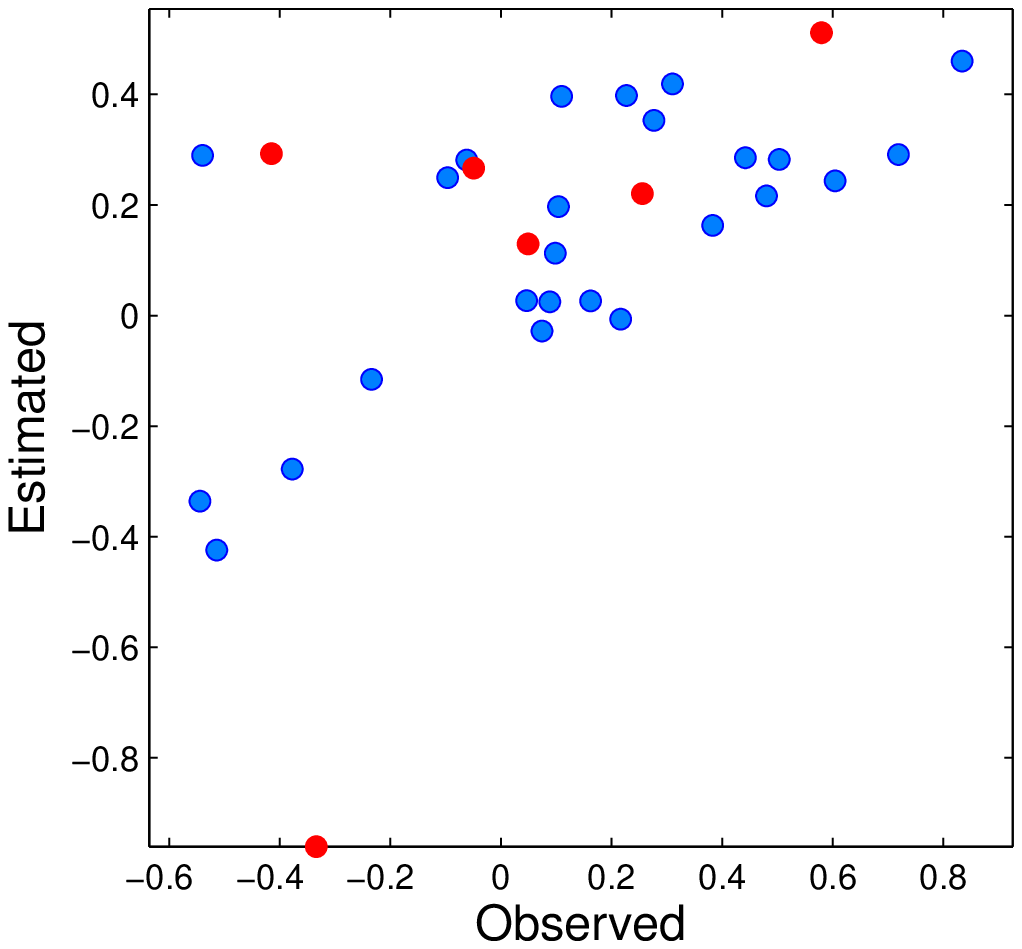}\\
    \includegraphics[scale=.4]{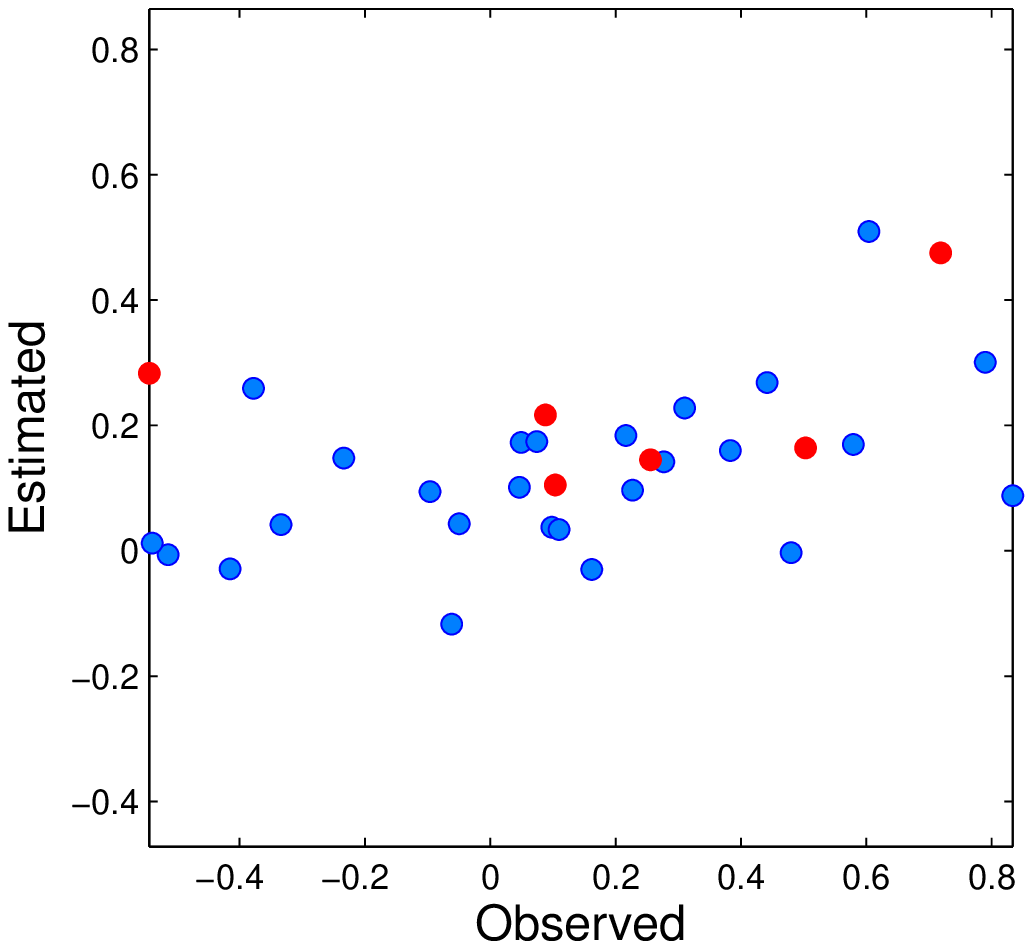}
    \includegraphics[scale=.4]{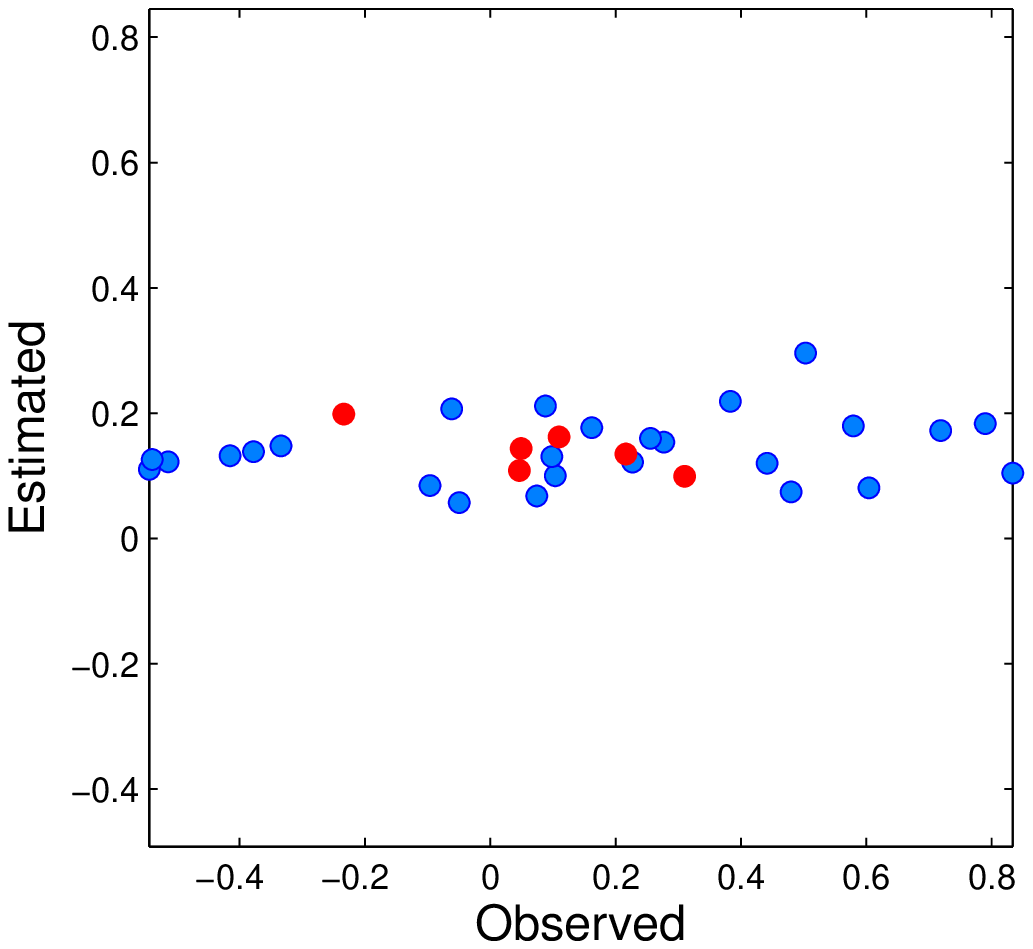}
    \includegraphics[scale=.4]{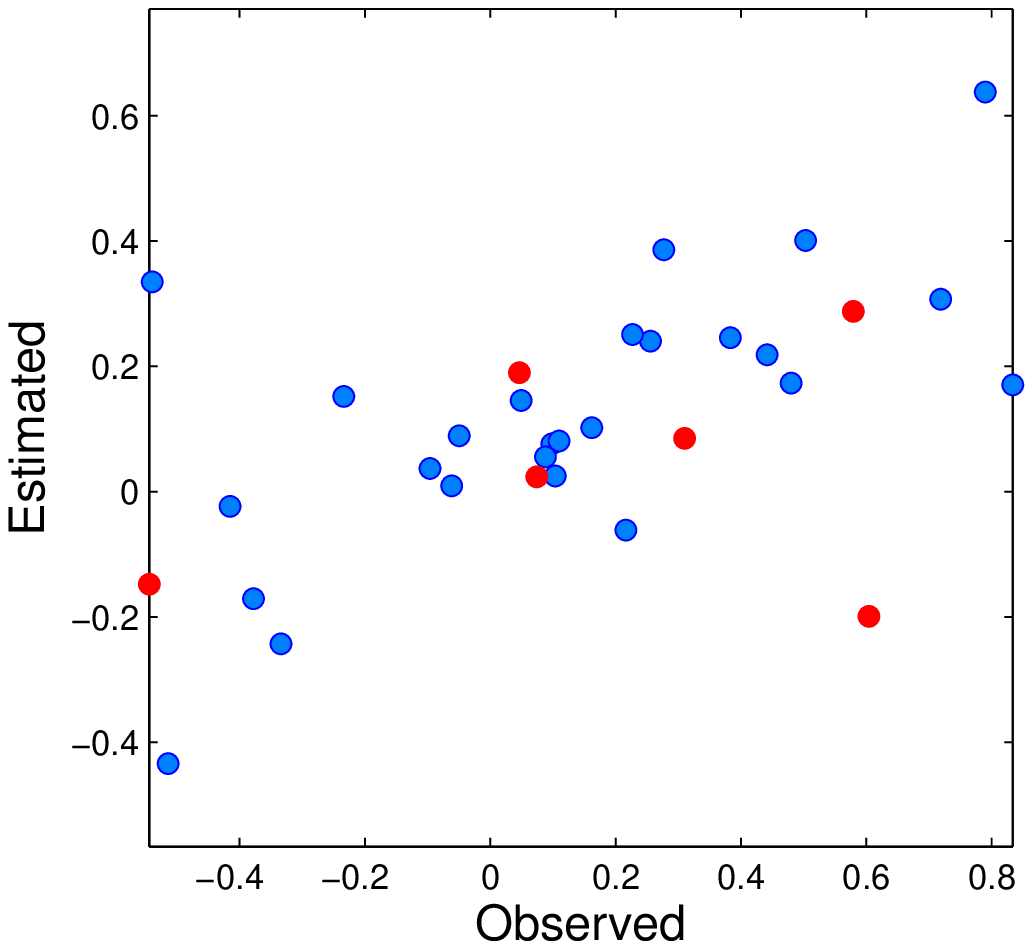}
    \caption{Normalized BMD estimation. The plots show examples of
      estimated versus observed normalized BMD for the training (blue) and
      prediction (red) sets. The six plots correspond to six random
      splittings of the data into training and prediction sets. }
    \label{fig:invivo:scatter}
\end{figure}

Figure~\ref{fig:invivo:scatter} shows some examples of
predictive performance of the regression
model~\eqref{eq:reg:mod:disc}. In each case, the sample is split
randomly into a training set and a test set, the latter
containing 2 rats from each of weeks 4,~6 and~8. We have also
discarded 5 rats from the sample as outliers, 
based on their average prediction error across all the
partitions -- for those five, the average prediction errors were
significantly higher than the rest. We refer to each such partition
of the data as a cross-validation or CV batch. Thus, each CV batch
contains 26 training and 6 test rats. A total of $50$ CV batches
were considered. For each batch, we have chosen the regularization
parameters to minimize the prediction error on the test set. We will call this the
\emph{adaptive} (or oracle) choice for the parameters. This gives the
best performance we could hope that the model achieves in each
case. It can be seen the results are mixed, even with this optimal
choice of the parameters, with some partitions of the data allowing
for a good prediction and some not.

\begin{figure}[t]
  \centering
  \begin{minipage}{0.45\textwidth}
    \centering
      \mbox{Adaptive Regularization}
      \includegraphics[scale=.6]{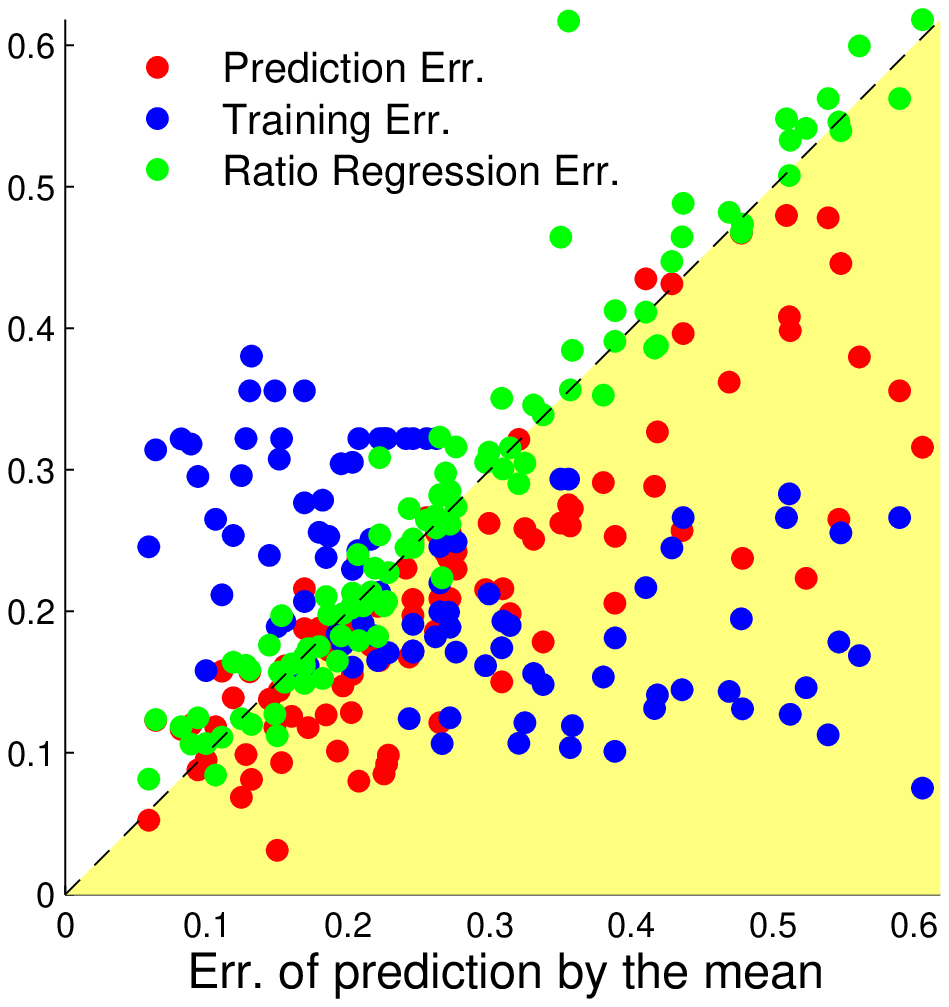}
  \end{minipage}
 \hspace{4ex}
 \begin{minipage}{0.45\textwidth}
      \centering
      \mbox{Fixed Regularization}
      \includegraphics[scale=.6]{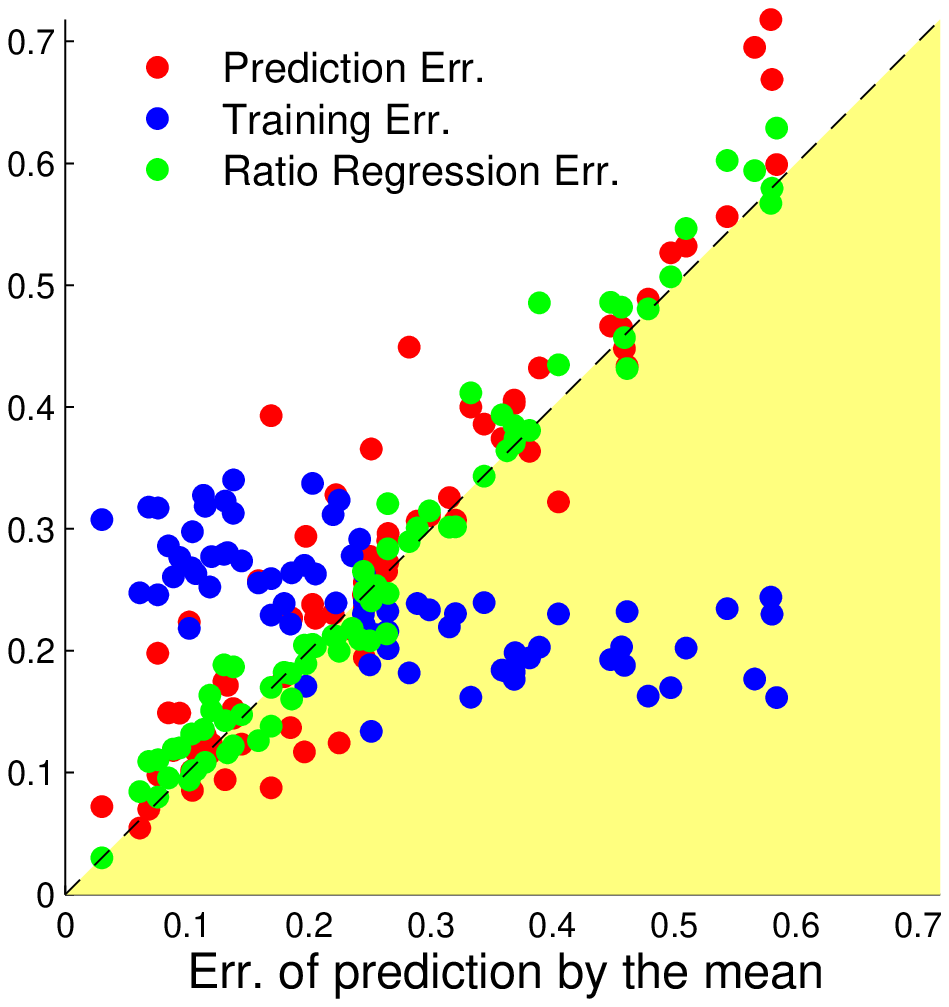}
  \end{minipage}
  \caption{Relative prediction performance. The plots show the mean
    absolute deviation error,
    in normalized BMD estimation, for the training (blue) and
    prediction (red) sets of our proposed regression
    model~\eqref{eq:reg:mod:disc}, versus the error of prediction by the
    mean.  Also shown (green) is the error of the simpler ratio
    regression approach. The two panels correspond to the adaptive
    versus fixed choice of regularization parameters in regression
    model~\eqref{eq:reg:mod:disc}.}
   \label{fig:rel:pred:perfor}
\end{figure}

The left panel of Figure~\ref{fig:rel:pred:perfor} illustrates the
same setting, but from the perspective of prediction errors. We have
used the median absolute deviation (MAD) as our measure of performance
(which will be the default throughout). The $x$-axis on this plot
shows the error of the \emph{prediction by the mean}, by which we mean
the error of the estimator that outputs the mean of the training set
(which is $\approx 0$) disregarding the Raman data. We will use the error
of this constant estimator as a baseline. The $y$-axis shows the
training error, and the test error of our model, together with
the test error of a simpler approach which we have called
\emph{ratio regression}, to be described shortly. Each point in this
plot corresponds to a CV batch. Each point below the diagonal line
shows a partition of the data that can be predicted better by the
model than the baseline.  With the adaptive choice of the
regularizer, this holds for almost all the batches.

The ratio regression approach predicts the BMD by regressing against a
simple spectral feature, namely the ratio of two peaks of interest, at
wavenumbers \SI{954}{cm^{-1}} and \SI{1450}{cm^{-1}}, which are known
to correspond to mineral (calcuim) and matrix (collagen),
respectively. The higher the ratio, the higher is the likelihood of
the bone being present in the specimen, as suspected by the
chemists.   The average of these two peaks (across all source-detector
positions) is computed for each rat, the mineral-to-matrix ratio is
computed by dividing the amplitude of \SI{954}{cm^{-1}} peak to that
of \SI{1450}{cm^{-1}}, and the result is used as the covariate to
predict the BMD.   This method, or a version of it, is typically used
by Raman spectroscopists.  Figure~\ref{fig:rel:pred:perfor} shows that
the result is very similar to the baseline of predicting by the mean,
meaning using Raman spectra in this fashion does not help predict BMD
in this experiment.

The right panel of Figure~\ref{fig:rel:pred:perfor} shows the results
for a more standard choice of the regularization parameters which we
have called \emph{fixed regularization}. In this case, a fixed set of
parameters is used for all batches, and they are chosen so that the
average error over all batches (i.e., the CV error) is minimized. With
this choice it can be seen that the performance of our model is
similar to that of ratio regression and the baseline).

\begin{figure}
  \centering
  \includegraphics[scale=0.45]{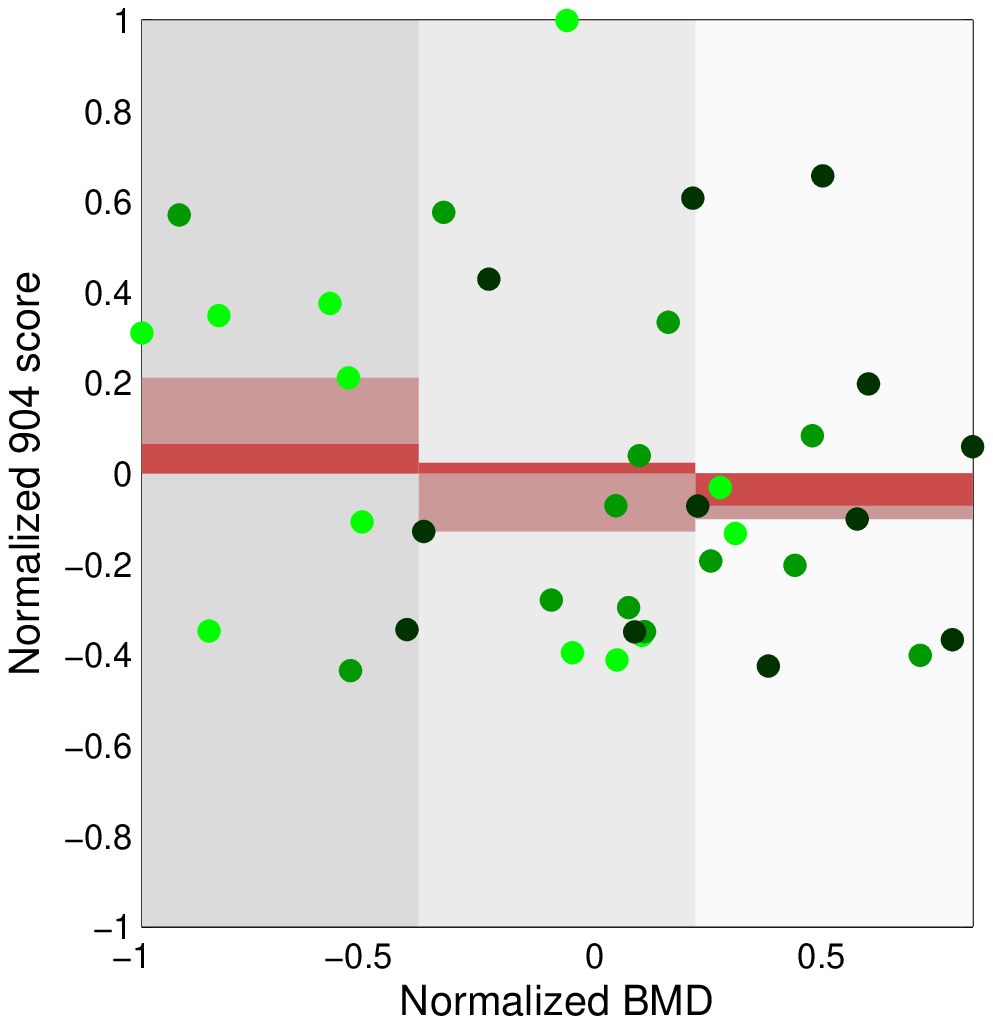}
  \includegraphics[scale=0.45]{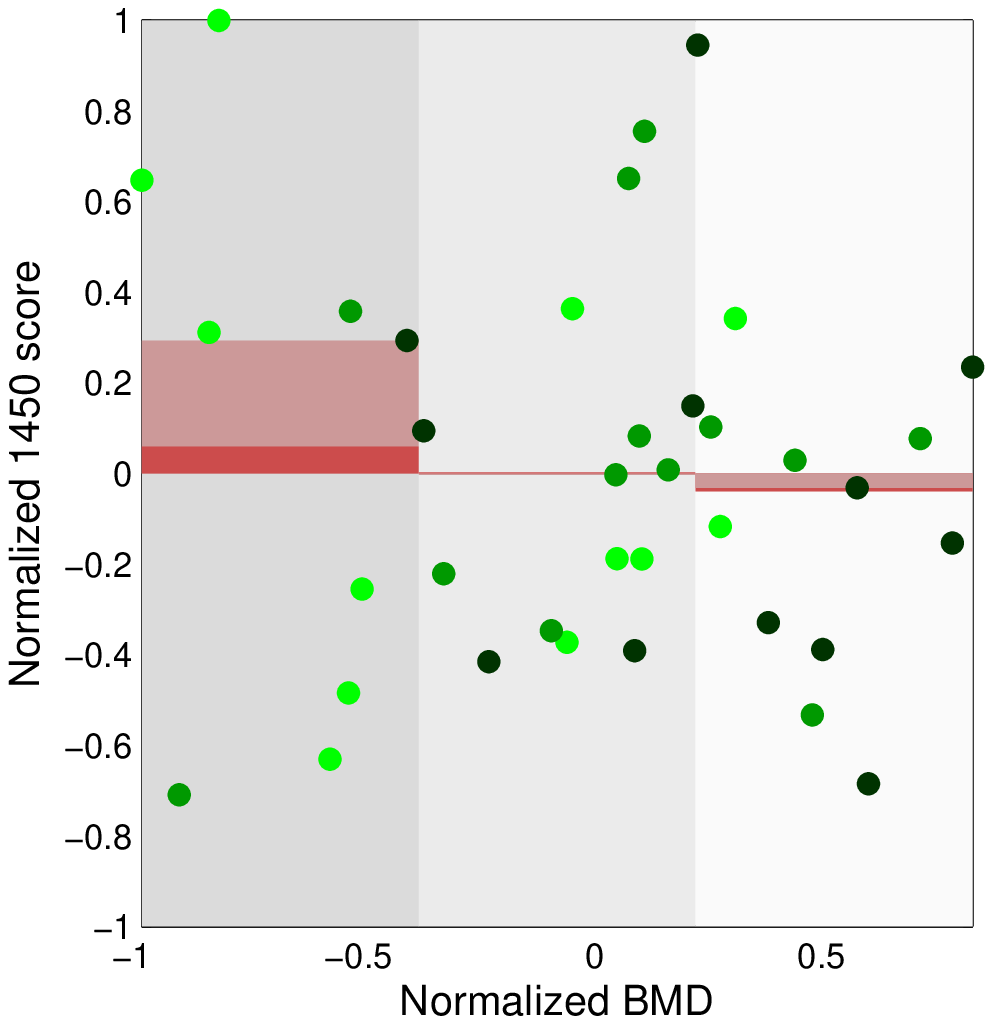}
  \includegraphics[scale=0.45]{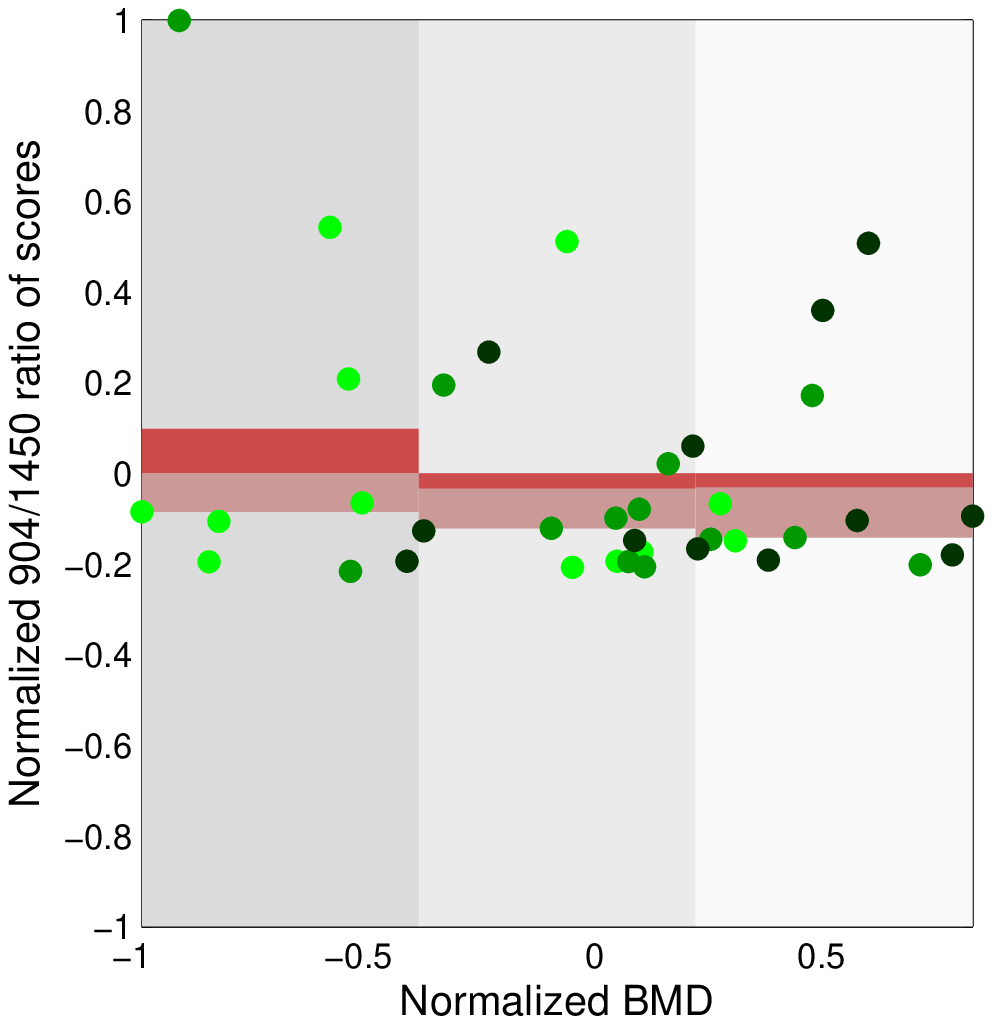}
  \caption{\small  Piecewise-constant approximation to regression curves for
    peaks of interest. Form left to right the plots show the
    normalized (and centered) scores obtained for
    \SI{954}{cm^{-1}} and \SI{1450}{cm^{-1}} wavenumbers and their ratio, versus the normalized
    BMD. The range of BMD is divided into three bands, and the mean
    (dark red) and the median (light red) of scores within each band
    are shown. The points are color-coded based on the week they
    belong to, with week 4, 6 and 8 being light, moderate and dark
    green, respectively.
    }
    \label{fig:piece:const:approx}
\end{figure}

We draw three main conclusions from these results. First, predicting BMD (or similar
measures) from in vivo Raman is inherently difficult, as is suggested
by the fact that both the simple ratio regression and our more
elaborate model with fixed regularization fail to show significant
improvement over the mean. This may be partly due to the mixing of
bone and tissue signals, partly due to variability in the rats and the
small sample size, and
partly due to the quality of the Raman data from this experiment.  
Second, the fact that we can successfully predict for some batches and not for the others
suggests that the rats are not homogeneous; if we had enough samples
from each (latent) group of similar rats, we might be able to improve
the prediction.  Finally, we suspect that the present dataset
might have been plagued by some calibration (or systematic) errors,
which precludes a global improvement on prediction performance. This
is suggested by the poor performance of the ratio regression, which
is believed to be a reasonably effective marker of
healing. Figure~\ref{fig:piece:const:approx} below also reinforces
this point.  In other words, only in situations where ratio regression
shows some improvement over the mean, we can hope and expect that
our method provides further improvements. The ex vivo results to be
discussed below support this claim.

\begin{figure}
  \centering
  \begin{minipage}{0.46\textwidth}
    \centering
    \mbox{Adaptive Regularization}\\
    \vspace{2ex}
      \includegraphics[scale=.5]{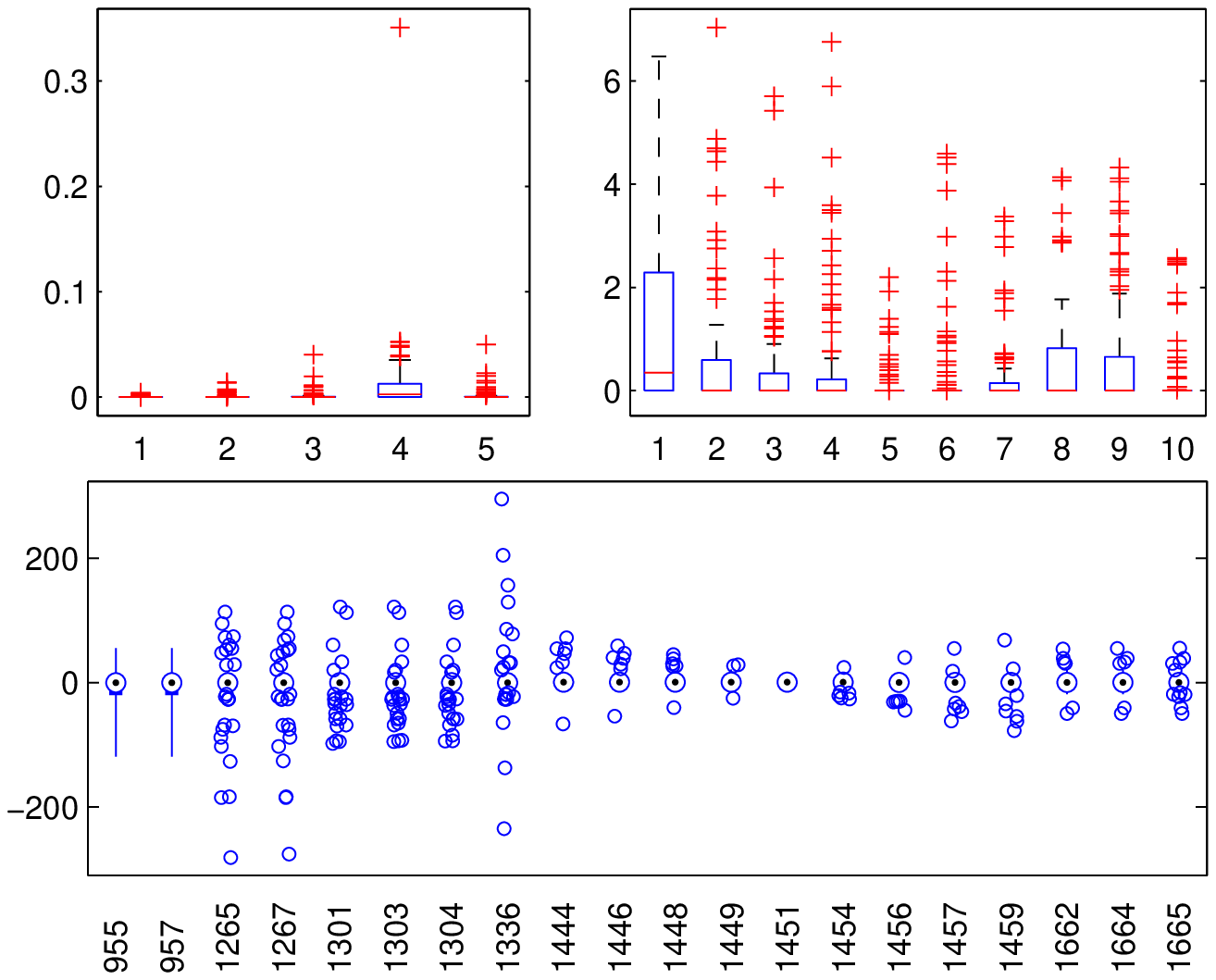}
  \end{minipage}
  \hspace{5ex}
  \begin{minipage}{0.46\textwidth}
    \centering
    \mbox{Fixed Regularization}
      \includegraphics[scale=.5]{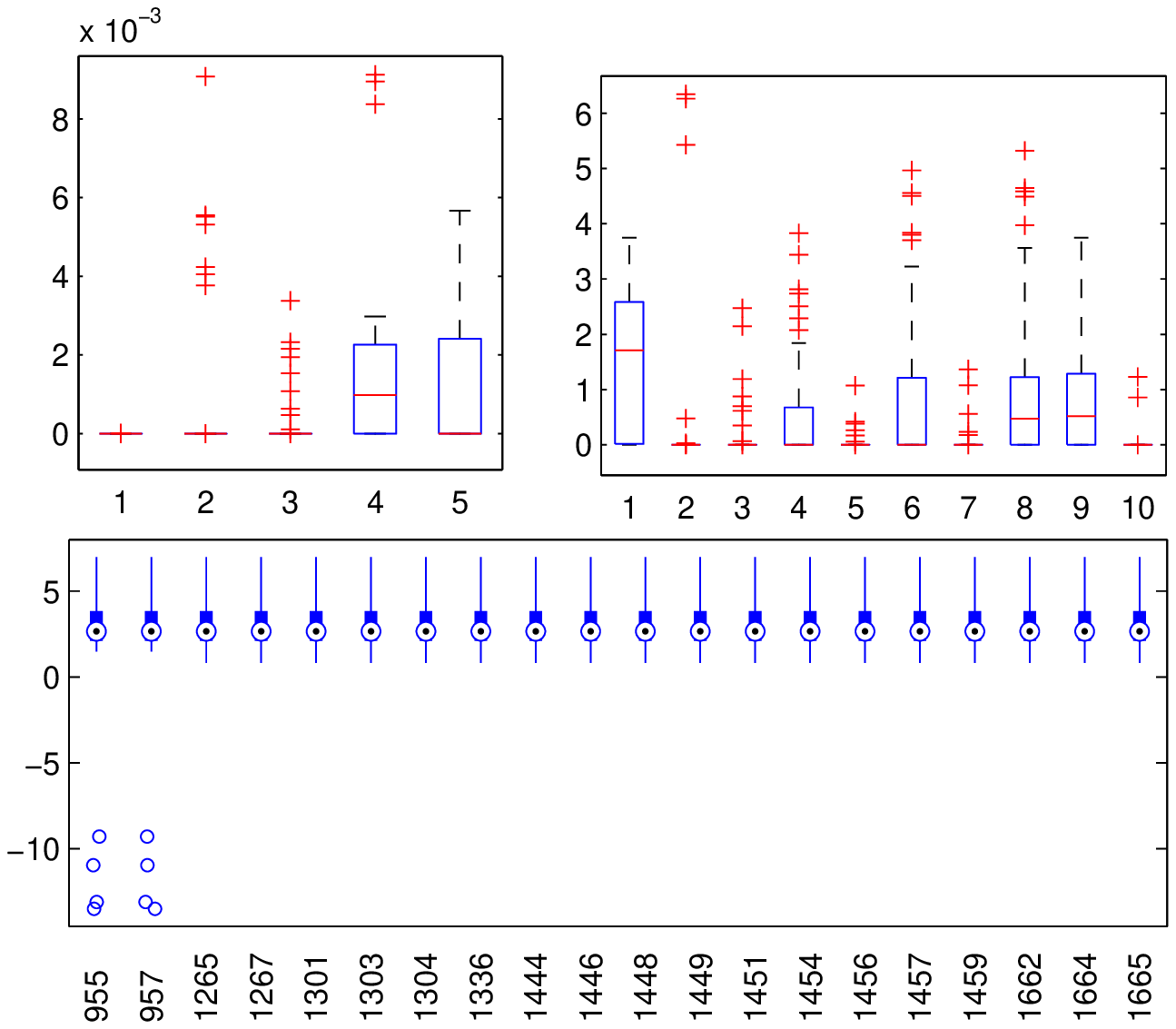}
  \end{minipage}
  \caption{\small Estimated model coefficients for the in vivo dataset, for
    both adaptive and fixed regularization. Boxplots of source weights
    ($5$-vector), detector weights ($10$-vector) and wavenumber
    weights ($40$-vector) are illustrated.}
  \label{fig:invivo:estim:coeff}
\end{figure}

\medskip Figure~\ref{fig:invivo:estim:coeff} shows the estimated
coefficients of the model.  The three sets of coefficients are the
$5$-vector of source weights ($\coa$), the $10$-vector of detector
weights ($\cob$) and the $40$-vector of wavenumber weights
($\coc$). The high number of outliers in source/detector weights and
the general tendency of the wavenumber weights to fluctuate around
zero are in alignment with conclusions above, that is,
no global correlation between this dataset and BMD can be inferred.

\medskip
We conclude this section with another look at the in vivo dataset from
the perspective of ratio
regression. Figure~\ref{fig:piece:const:approx} illustrates scatter
plots for the two peaks of interest, at $954$ and
$\SI{1450}{cm^{-1}}$, and their ratios, versus the (normalized)
BMD. The values for the peaks are those obtained from our functional
representation~\footnote{The same qualitative behavior is observed if
  we use the raw peaks, but they show higher variability. This is due
  to the shrinkage effect of our representation.}, and are centered
and normalized to maximum absolute value $1$.

The range of BMD is divided into three bands and the mean and median
of the $y$-axis data is shown in each band. This provides a
piecewise-constant approximation to the regression curve. These plots
show the somewhat anomalous nature of the dataset. We would expect
that as ossification progresses (BMD increases) the calcium-related
\SI{954}{cm^{-1}} should increase, while the collagen-related
\SI{1450}{cm^{-1}} should decrease. A near reversal of this trend is
observed for the mean behavior of \SI{954}{cm^{-1}}, while the
\SI{1450}{cm^{-1}} follows the expected relationship to some
extent. As a result, their ratio remains almost constant, on average,
over the entire range of BMD (with a slight deviation towards the
lower end), preventing a possible prediction. This further reinforces
our suspicion of a calibration error; getting the ring aligned with
exact location of the defect is, in general, difficult. The current
version of the ring also was not in contact with the leg. Thus, slight
variations in its position might have caused nearby healthy bone to
contribute to the spectra. (More precise construction of the ring has
been proposed and is under investigation.)

\subsection{Ex vivo Raman}\label{sec:results:exvivo}
For ex vivo data of Section~\ref{sec:exvivo:Raman:data}, there is no
spatial dynamics to be modeled. That is, $p = d = 1$, and the model
reduces to a more or less classical functional regression with Lasso
and fused Lasso penalties. In this case, we only have one set of
parameters to estimate, namely, $\{\coc_\fvarp\}$ or equivalently, the
functional weights $\coC(t)$ of~\eqref{eq:func:weights}.

\begin{figure}
  \begin{minipage}{.35\linewidth}
  \centering
  Cross-validation Errors\\[1ex]
     \includegraphics[scale = 0.5]{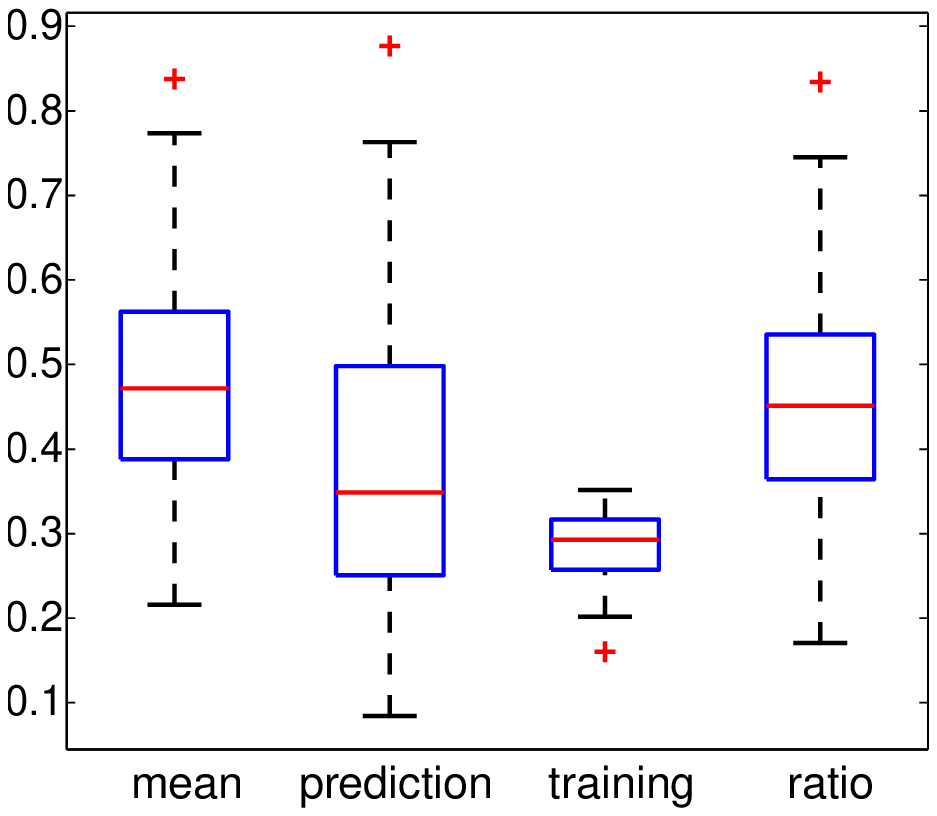}
  \end{minipage}
  \begin{minipage}{.6\linewidth}
  \centering
    Estimated Coefficients\\[1ex]
    \includegraphics[scale = 0.6]{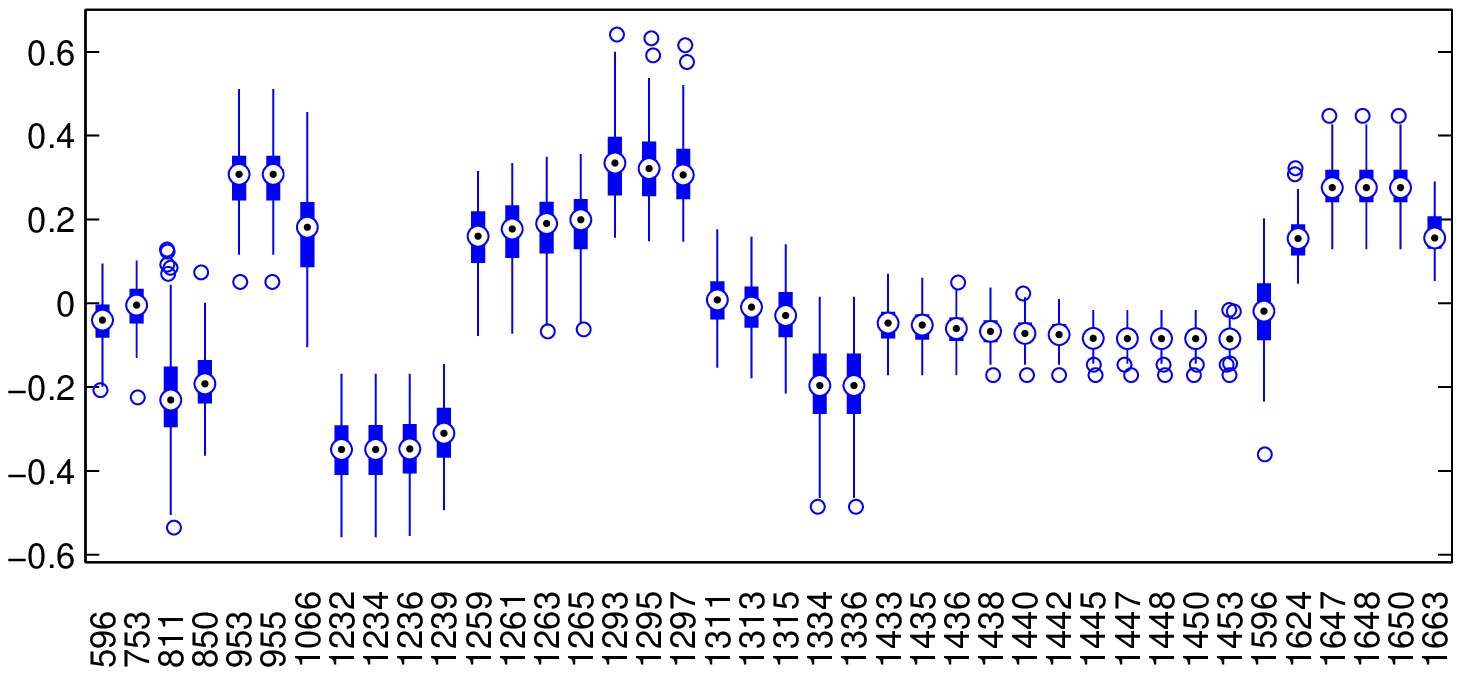}
  \end{minipage}
  \label{fig:ex-vivo:results}
  \caption{\small The results for ex vivo dataset. Left panel shows, from
    left to right, CV errors
  for the baseline (prediction by the mean), prediction error for our
  model followed by its training error, and the error of ratio
  regression. Right panel shows the estimated model coefficients.}
\end{figure}

Figure~\ref{fig:ex-vivo:results} (left panel) shows cross-validation
(CV) errors for our model (in both prediction and training) versus the
baseline (prediction by the mean) and ratio regression. The latter two
are as described in the context of in vivo data
(cf. Section~\ref{sec:results:invivo}). Here, the parameters of our
model are set by CV in the usual sense; that is, a fixed set of
parameters are used for all the batches, chosen so that average CV
error is minimized. The partitioning of data into prediction and
training sets, and the number of batches are as in
Section~\ref{sec:results:invivo}. One observes that our model provides
a fair improvement (on average) over the simpler ratio regression
which is only slightly better than the baseline.

Also shown in Figure~\ref{fig:ex-vivo:results} (right panel) are the
boxplots of the estimated coefficients $\{\coc_\fvarp\}$. In contrast
to the in vivo case, one can clearly identify regions (wavenumbers)
that consistently exhibit large coefficients and hence are helpful in 
predicting the BMD. For example, the mineral band
at $\approx 954$ (which is known to be excited by calcium in the
bone) is prominent in the plot. Another interesting observation is
that of the two collagen bands, one at $\approx \SI{1450}{cm^{-1}}$
and the other at $\approx \SI{1660}{cm^{-1}}$, the latter seems to be
more correlated with BMD, despite the fact that $\SI{1450}{cm^{-1}}$
is in general the highest amplitude peak in the Raman spectra
(cf. Figure~\ref{fig:func:rep}).   Overall, these results suggest that
the main challenges in using Raman spectra to predict BMD come from
the attempt to do it in vivo, rather than some inherent problem with
the bone spectra themselves.

\subsection{NMR}

The regression model introduced in~\ref{sec:reg:model} has two spatial
dimensions, which is one more than what is needed for the NMR dataset;
we discard one dimension by setting, say, $d = 1$. Then, the model has
two sets of parameters to be estimated, $\{\coa_i\}$ and
$\{\coc_\fvarp\}$. As with the Raman datasets, we retain the $N = 40$
highest scoring chemical shifts in the model.

\begin{figure}
  \begin{minipage}{.35\linewidth}
  \centering
   Cross-validation Errors
  \includegraphics[scale = 0.5]{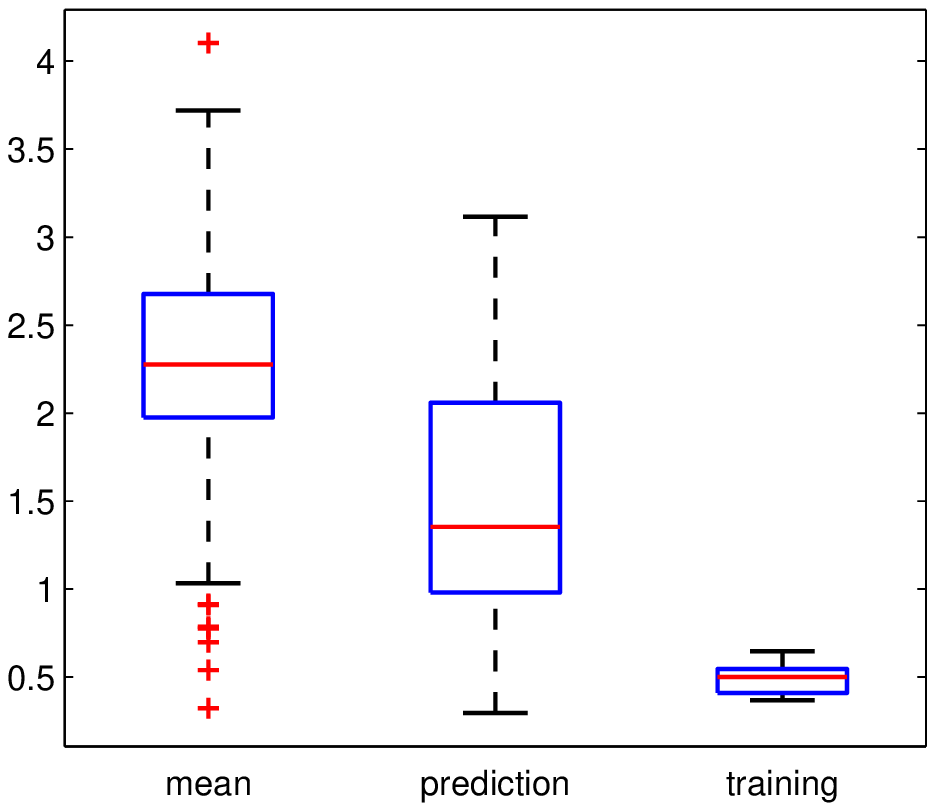}
  \end{minipage}
  \begin{minipage}{.6\linewidth}
     \centering
       Estimated Coefficients ($\alpha,\gamma$)
    \includegraphics[scale = 0.53]{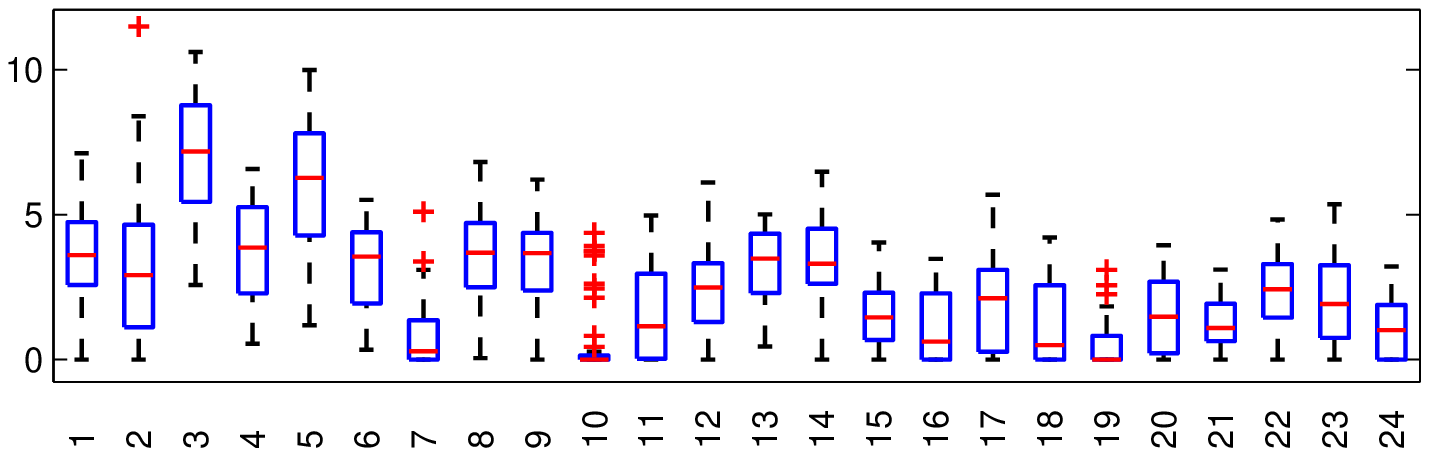}\\
    \includegraphics[scale = 0.53]{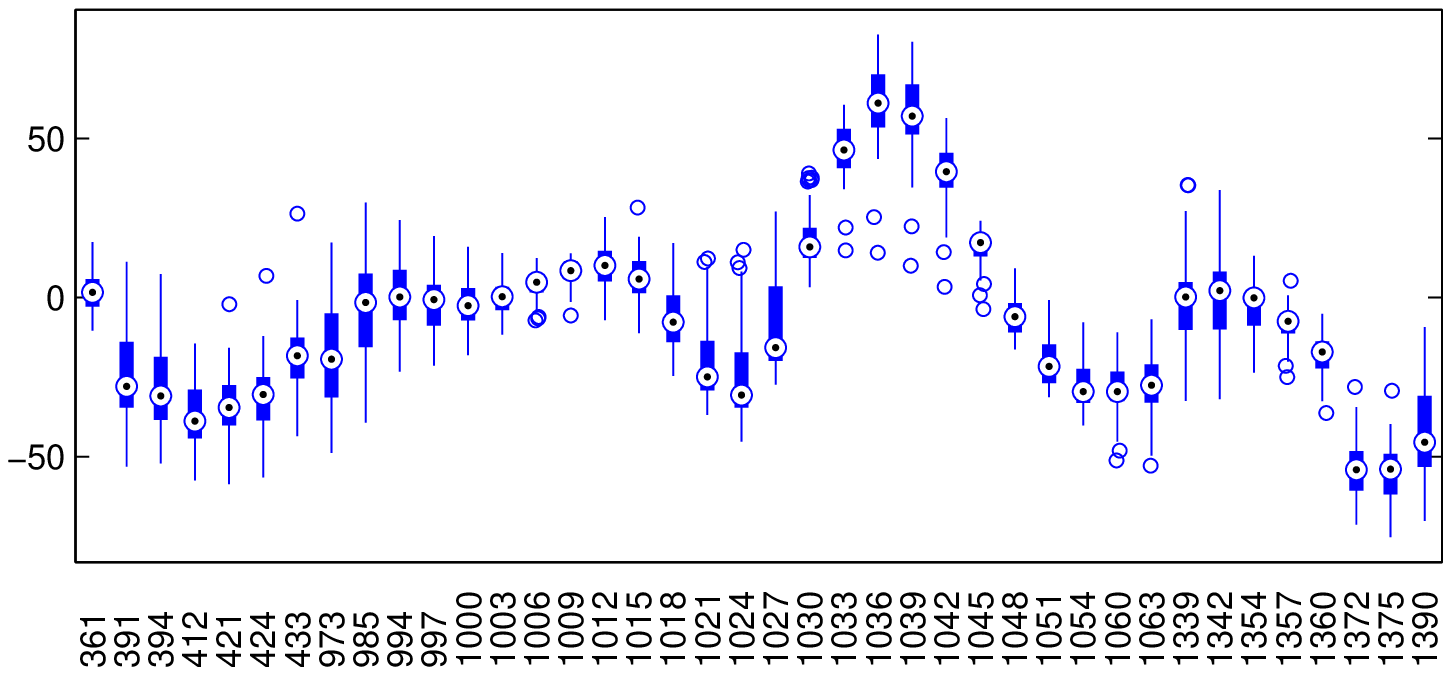}
  \end{minipage}
  \caption{The results for NMR dataset.  Left panel shows, from
    left to right, CV errors
  for the baseline (prediction by the mean), and prediction error for our
  model followed by its training error.  Right panel shows the estimated model coefficients.}
  \label{fig:nmr:results}
\end{figure}

Figure~\ref{fig:nmr:results} shows cross-validation (CV) results for
the NMR dataset. A total of 50 batches were used for CV, where in each
batch we left 4 samples out for prediction (out of a total of
$n=20$). The left panel in the figure shows CV errors for the baseline
(prediction by the mean) and for our model (both prediction and
training errors). We observe a substantial improvement by our model
over the baseline.  The left panel shows the boxplots for estimated
coefficients $\alpha$ (on the top) and $\gamma$ (on the bottom). These
clearly show some spatial and spectral positions to predict the
response. It further illustrates that with a sufficiently high
signal-to-noise ratio, our models can be used for prediction and
variable selection.

\begin{figure}

\end{figure}


\section{Discussion}\label{sec:discuss}

We presented a functional model for spectra which can be used for
denoising and compression. Based on this representation of the data,
we proposed a regression model to predict a scalar response (e.g.,
bone mineral density or lipoprotein concentration) based on
multi-dimensional spectroscopy data. The data was modeled as a tensor
with several spatial dimensions and one spectral dimension. A rank-one
multilinear map was proposed to describe the relation between the
spectra and the response, with some structure on the coefficients,
namely sparsity in the spectral domain, and dependence based on a
similarity measure.  The structure was enforced using regularization,
with $\ell_1$ penalty to induce sparsity, and a fused Lasso type
penalty to enforce similarity. (Some structure was also assumed for
the spatial coefficients, albeit a minimal one.)

We considered the effectiveness of the approach in three settings: in
vivo and ex vivo Raman data from a fracture healing experiment, and
using NMR data in a lipoprotein concentration study. For the in vivo
Raman experiments, our results were mixed. It was possible to predict
the outcome for some partitions of the data into training and
prediction sets, but not for others. Overall our model and a simpler
ratio regression model suggested that there was no significant global
relation between the in vivo Raman data at hand and the BMD
outcome. We conjectured three possible causes: (1) The problem of
prediction based on in vivo Raman data is hard in general, due to
bone-tissue mixing, low signal-to-noise ratio and normalization
issues. (2) The prediction is possible for a more homogeneous group of
rats than what we had. This was suggested by good results if we used
adaptive regularization (i.e., picked the best choice for each batch
of CV). (3) The data at hand may have been corrupted by
calibration and systematic errors. This was further corroborated by
looking at piecewise constant approximations to regression curve for
predicting BMD based on single peaks in the spectra, which are known
to correlate with bone composition. This issue is partly due to the
difficulty of aligning the measuring device with the exact location of
the fracture, which is a known challenge that is under
investigation. For the ex vivo Raman data, and the NMR, we showed that
the model has (global) predictive power, and can improve significantly
over the baseline of prediction by the mean.

In this paper we mainly focused on regression models. Another
interesting possibility for future work is to look at PCA type
analyses, by which we mean models taking into account a decomposition
of the spectra into relevant (and irrelevant) components. A promising
direction is to use the response variable to guide the selection of
components. Another feature of the problem, namely the positivity of
spectra, can also be taken into account, which makes the problem
different from classical PCA, or PCA-based regression. A challenge in
this case is to find alternatives to orthogonality which are
meaningful for positive components.


\section*{Acknowledgements}
This research was supported by NIH grant 5-R01-AR-056646-03.  We are grateful to Michael Morris (Chemistry, University of Michigan) and members of his lab, especially Paul Okagbare and Francis Esmonde-White, as well as Steven Goldstein (University of Michigan, Orthopedic Surgery) and members of his lab for collaborating on the grant and collecting and sharing the data, as well as for many useful discussions.

\bibliographystyle{plain}
\bibliography{raman-refs2}

\begin{thebibliography}{10}

\bibitem{Berlinet2004}
A~Berlinet and C~Thomas-Agnan.
\newblock {\em {Reproducing Kernel Hilbert Spaces in Probability and
  Statistics}}.
\newblock Springer US, Boston, MA, 2004.

\bibitem{Chen2012}
X~Chen, Q~Lin, and S~Kim.
\newblock {Smoothing proximal gradient method for general structured sparse
  regression}.
\newblock {\em The Annals of Applied Statistics}, 6(2):719--752, June 2012.

\bibitem{Chew2002}
W~Chew, E~Widjaja, and M~Garland.
\newblock {Band-Target Entropy Minimization (BTEM): An Advanced Method for
  Recovering Unknown Pure Component Spectra. Application to the FTIR Spectra of
  Unstable Organometallic Mixtures}.
\newblock {\em Organometallics}, 21(9):1982--1990, April 2002.

\bibitem{Dyrby2005}
M~Dyrby, M~Petersen, A~K Whittaker, L~Lambert, L~N{\"o}rgaard, Rasmus Bro, and
  S\o ren~Balling Engelsen.
\newblock {Analysis of lipoproteins using 2D diffusion-edited NMR spectroscopy
  and multi-way chemometrics}.
\newblock {\em Analytica Chimica Acta}, 531(2):209--216, February 2005.

\bibitem{Hanlon2000}
E~B Hanlon, R~Manoharan, T~W Koo, K~E Shafer, J~T Motz, M~Fitzmaurice, J~R
  Kramer, I~Itzkan, R~R Dasari, and M~S Feld.
\newblock {Prospects for in vivo Raman spectroscopy.}
\newblock {\em Physics in medicine and biology}, 45(2):R1--59, February 2000.

\bibitem{Koltchinskii2011}
V~Koltchinskii, K~Lounici, and A~B Tsybakov.
\newblock {Nuclear-norm penalization and optimal rates for noisy low-rank
  matrix completion}.
\newblock {\em The Annals of Statistics}, 39(5):2302--2329, October 2011.

\bibitem{Maher2013}
J~R Maher, J~A Inzana, H~A Awad, and A~J Berger.
\newblock {Overconstrained library-based fitting method reveals age- and
  disease-related differences in transcutaneous Raman spectra of murine bones.}
\newblock {\em Journal of biomedical optics}, 18(7):077001, July 2013.

\bibitem{Martin2004}
A~A Martin, R~A {Bitar Carter}, L~{de Oliveira Nunes}, E~A {Loschiavo Arisawa},
  and L~{Silveira, Jr.}
\newblock {Principal components analysis of FT-Raman spectra of ex vivo basal
  cell carcinoma}.
\newblock In {\em SPIE 5321, Biomedical Vibrational Spectroscopy and Biohazard
  Detection Technologies}, volume 5321, pages 198--204, July 2004.

\bibitem{Matousek2006}
P~Matousek, E~R~C Draper, A~E Goodship, I~P Clark, K~L Ronayne, and Anthony~W
  Parker.
\newblock {Noninvasive Raman spectroscopy of human tissue in vivo.}
\newblock {\em Applied spectroscopy}, 60(7):758--63, July 2006.

\bibitem{Meier2005}
R~J Meier.
\newblock {On art and science in curve-fitting vibrational spectra}.
\newblock {\em Vibrational Spectroscopy}, 39(2):266--269, October 2005.

\bibitem{Negahban2011}
S~Negahban and M~J Wainwright.
\newblock {Estimation of (near) low-rank matrices with noise and
  high-dimensional scaling}.
\newblock {\em The Annals of Statistics}, 39(2):1069--1097, April 2011.

\bibitem{Pauca2006}
V~P Pauca, J~Piper, and R~J Plemmons.
\newblock {Nonnegative matrix factorization for spectral data analysis}.
\newblock {\em Linear Algebra and its Applications}, 416(1):29--47, July 2006.

\bibitem{Scholkopf2001}
B~Sch\"{o}lkopf, R~Herbrich, and A~J Smola.
\newblock {A generalized representer theorem}.
\newblock {\em Computational learning theory}, pages 416--426, 2001.

\bibitem{Tchanque-Fossuo2013}
C~N Tchanque-Fossuo, B~Gong, B~Poushanchi, A~Donneys, D~Sarhaddi, K~Kelly
  Gallagher, Sagar~S Deshpande, Steven~a Goldstein, Michael~D Morris, and
  Steven~R Buchman.
\newblock {Raman spectroscopy demonstrates Amifostine induced preservation of
  bone mineralization patterns in the irradiated murine mandible.}
\newblock {\em Bone}, 52(2):712--7, February 2013.

\bibitem{Tibshirani1996}
R~Tibshirani.
\newblock {Regression shrinkage and selection via the lasso}.
\newblock {\em Journal of the Royal Statistical Society. Series B},
  58(1):267--288, 1996.

\bibitem{Tibshirani2005}
R~Tibshirani, M~Saunders, S~Rosset, J~Zhu, and K~Knight.
\newblock {Sparsity and smoothness via the fused lasso}.
\newblock {\em Journal of the Royal Statistical Society: Series B},
  67(1):91--108, February 2005.

\bibitem{Vrabie2007}
V~Vrabie, C~Gobinet, O~Piot, A~Tfayli, P~Bernard, R\'{e}gis Huez, and Michel
  Manfait.
\newblock {Independent component analysis of Raman spectra: Application on
  paraffin-embedded skin biopsies}.
\newblock {\em Biomedical Signal Processing and Control}, 2(1):40--50, January
  2007.

\bibitem{Widjaja2003}
Effendi Widjaja, Nicole Crane, Tso-Ching Chen, Michael~D Morris, Michael~A
  Ignelzi, and Barbara~R McCreadie.
\newblock {Band-target entropy minimization (BTEM) applied to hyperspectral
  Raman image data.}
\newblock {\em Applied spectroscopy}, 57(11):1353--62, November 2003.

\bibitem{Zhao2013}
J~Zhao and C~Leng.
\newblock {Structured Lasso for Regression with Matrix Covariates}.
\newblock {\em Statistica Sinica}, To appear, 2013.

\end{thebibliography}

\end{document}